\documentclass[3p,authoryear]{elsarticle}

\journal{Icarus}

\usepackage{natbib}

\usepackage{doi}
\usepackage{amsmath}
\usepackage{textcomp}
\usepackage{color}
\usepackage[usenames,dvipsnames]{xcolor}
\usepackage{ulem}
\usepackage{hyperref}
\hypersetup{colorlinks=true, urlcolor=blue}

\usepackage{lineno}

    \newcommand*\patchAmsMathEnvironmentForLineno[1]{%
      \expandafter\let\csname old#1\expandafter\endcsname\csname #1\endcsname
      \expandafter\let\csname oldend#1\expandafter\endcsname\csname end#1\endcsname
      \renewenvironment{#1}%
         {\linenomath\csname old#1\endcsname}%
         {\csname oldend#1\endcsname\endlinenomath}}%
    \newcommand*\patchBothAmsMathEnvironmentsForLineno[1]{%
      \patchAmsMathEnvironmentForLineno{#1}%
      \patchAmsMathEnvironmentForLineno{#1*}}%
    \AtBeginDocument{%
    \patchBothAmsMathEnvironmentsForLineno{equation}%
    \patchBothAmsMathEnvironmentsForLineno{align}%
    \patchBothAmsMathEnvironmentsForLineno{flalign}%
    \patchBothAmsMathEnvironmentsForLineno{alignat}%
    \patchBothAmsMathEnvironmentsForLineno{gather}%
    \patchBothAmsMathEnvironmentsForLineno{multline}%
    }

\mathchardef\mhyphen="2D

\begin{document}

\begin{frontmatter}



\title{In-flight calibration of the Dawn Framing Camera II: Flat fields and stray light correction\tnoteref{label1}\tnoteref{label2}}
\tnotetext[label1]{\doi{10.1016/j.icarus.2014.02.018}}
\tnotetext[label2]{\copyright 2017. This manuscript version is made available under the CC-BY-NC-ND 4.0 licence:\\ \url{https://creativecommons.org/licenses/by-nc-nd/4.0/}}


\author[DLR]{S.E.~Schr\"oder}
\author[DLR]{S.~Mottola}
\author[DLR]{K.-D.~Matz}
\author[DLR]{T. Roatsch}

\address[DLR]{Deutsches Zentrum f\"ur Luft- und Raumfahrt (DLR), 12489 Berlin, Germany}

\begin{abstract}

The NASA Dawn spacecraft acquired thousands of images of asteroid Vesta during its year-long orbital tour, and is now on its way to asteroid Ceres. A method for calibrating images acquired by the onboard Framing Camera was described by Schr\"oder et al.\ (2013; Icarus 226, 1304). However, their method is only valid for point sources. In this paper we extend the calibration to images of extended sources like Vesta. For this, we devise a first-order correction for in-field stray light, which is known to plague images taken through the narrow band filters, and revise the flat fields that were acquired in an integrating sphere before launch. We used calibrated images of the Vesta surface to construct simple photometric models for all filters, that allow us to study how the spectrum changes with increasing phase angle (phase reddening). In combination with these models, our calibration method can be used to create near-seamless mosaics that are radiometrically accurate to a few percent. Such mosaics are provided in JVesta, the Vesta version of the JMARS geographic information system.

\end{abstract}


\end{frontmatter}



\section{Introduction}

The NASA Dawn spacecraft left asteroid Vesta in August 2012 after a successful science campaign, and is on its way to rendezvous with asteroid Ceres in April 2015. A method for calibrating images of both models of the onboard Framing Camera (FC1 and FC2) based on in-flight observations was outlined by \citet{S13a}. However, their calibration is only valid for point sources. To extend the method to resolved images of Vesta, it is necessary to characterize and correct for in-field stray light that is known to affect FC narrow band images of extended sources \citep{Si11,S13a}. It was realized after the launch of Dawn that in-field stray light also affects the images of the inside of the integrating sphere from which the flat fields were derived. It it therefore necessary to revise these flat fields as well. Such a revision also benefits from a comprehensive analysis of the thousands of images of Vesta acquired through each filter. Removing the stray light according to our method and using the revised flat fields allows for calibrating narrow-band FC images with a radiometric accuracy of a few percent, depending on the scene. The availability of fully calibrated images for several imaging campaigns at Vesta allows us to derive photometric models for each filter, which can be used to construct seamless mosaics of the surface. To allow the user to calibrate the images him- or herself, we archived our revised flat fields and stray light patterns in the Dawn Public Data archive\footnote{\url{http://dawndata.igpp.ucla.edu}}. Mosaics constructed from images calibrated with the method in this paper in combination with \citet{S13a} are also provided in the archive, and in JVesta, the Vesta version of the public geographic information system JMARS\footnote{\url{http://jmars.asu.edu}} \citep{C09}. Note that in this paper we display the FC image identification numbers in {\bf bold}, and Vesta observational campaign names in {\it italics}.

\section{Extending the Calibration}

\subsection{Calibration pipeline}

The calibration by \citet{S13a} is valid for images of point sources. To extend the calibration to extended sources like Vesta requires knowledge of the flat fields and the in-field stray light that is known to affect the narrow band filter images \citep{Si11}. First, we summarize the workings of the calibration pipeline.

The spectral radiance $\mathbf{L}^i$ (in W m$^{-2}$ nm$^{-1}$ sr$^{-1}$) of an extended source observed by prime camera FC2 through filter $i$ is retrieved from the raw image $\mathbf{A}^i$ (in DN) as
\begin{equation}
\mathbf{L}^i = \frac{\mathbf{C}^i}{\mathcal{R}^i \mathbf{N}^i} = \frac{(\mathbf{A}^i - \mathbf{S} - b)/t_{\rm exp} - \mathbf{D} - \mathbf{I}^i}{\mathcal{R}^i \mathbf{N}^i},
\label{eq:cal_pipeline}
\end{equation}
where $\mathbf{C}^i$ is the clean image (in DN~s$^{-1}$), $\mathcal{R}^i$ is the responsivity factor (in J$^{-1}$ m$^2$ nm sr), $\mathbf{N}^i$ is the normalized flat field (dimensionless), $\mathbf{S}$ is the smear image (in DN), $b$ is the bias (in DN), $t_{\rm exp}$ is the exposure time (in s), $\mathbf{D}$ is the dark current image (in DN s$^{-1}$), and $\mathbf{I}^i$ is the in-field stray light image (in DN s$^{-1}$). Based on the camera design, we may assume that the out-of-field stray light contribution is negligible. Residual charge, present on the CCD at the start of an exposure \citep{S13a}, has not been detected for FC2. For backup camera FC1, the calculation of the spectral radiance is as in Eq.~\ref{eq:cal_pipeline}, but, in addition, involves the subtraction of residual charge. We assume that the FC1 stray light patterns are identical to those of FC2 (which is the assumption behind the reconstruction of the patterns, as explained in Sec.~\ref{sec:stray_light}). The flat fields are unique to each camera. Prior to launch, FC images were obtained of the inside of an integrating sphere, and it is natural to assume that these images may serve as the flat fields $\mathbf{N}^i$. However, this approach is not recommended as we will now explain.

\subsection{Laboratory flat fields}
\label{sec:lab_flats}

\begin{figure}
\centering
\includegraphics[width=9cm,angle=0]{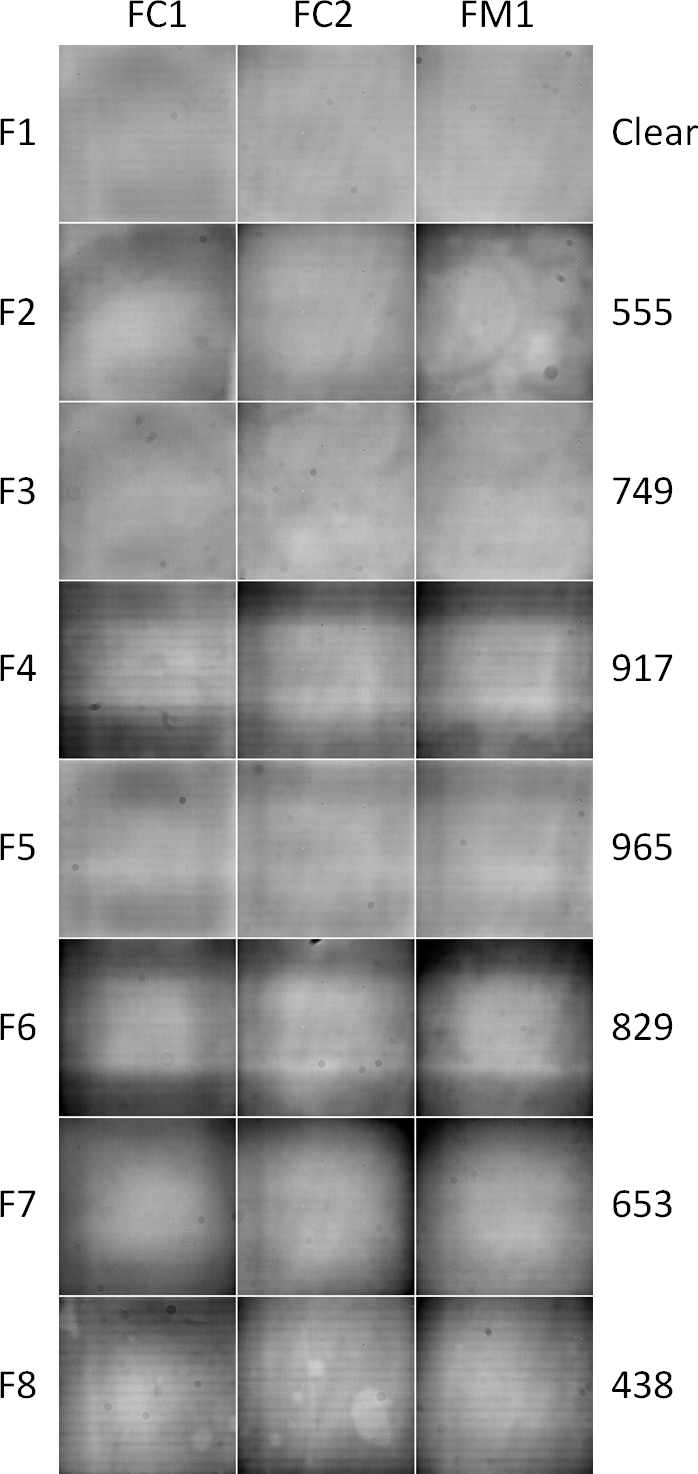}
\caption{Images of the inside of an integrating sphere acquired by, from left to right, FC1, FC2, and FM1, through each of the filters (number and effective wavelength in nm indicated). Flat fields are normalized to the same spot in the center, and displayed with black and white being 0.90 and 1.05, respectively.}
\label{fig:FC_flats}
\end{figure}

Images were obtained of the inside of an integrating sphere with the FC in thermal vacuum. We refer to these images as ``lab flat fields''. We display the FC1 and FC2 lab flat fields \citep{Si11} in Fig.~\ref{fig:FC_flats}, in combination with those acquired by the first flight model (FM1). The second flight model (FM2) and first flight spare (FS1) are now known as FC2 and FC1, respectively. The figure shows that the lab flat fields are decidedly non-flat. This is partly due to factors that are traditionally associated with flat fields, like pixel-to-pixel responsivity variations, dust particles on and inhomogeneities in the filter, and vignetting. The pixel-to-pixel variations are associated with responsivity variations across the CCD, including those due to (partial) obstruction of light by particles on its surface. Evidence for vignetting is seen especially well in the images taken through filter F7. The darkest corner in the FC2 image is the top right one, whereas in the FM1 image it is the top left one. The opposite corner is not appreciably darkened. Apparently, darkening (vignetting) critically depends on the position of the filter in the filter wheel socket, where the FC1 F7 filter appears well centered. Dust on the filters shows up as small dark disks in almost all flat fields, whereas filter inhomogeneities appear as large bright blobs in the FC2 F8 image.

Some of the non-flatness of the laboratory flat fields could be associated with the experiment or be specific to the optics of this camera, with factors like:
\begin{itemize}
\item inhomogenous illumination
\item reflectance inhomogeneities on the inside of the sphere
\item in-field-stray light
\item out-of-field-stray light
\item stray light from experimental setup
\end{itemize}
Disentangling the effects of these factors is not an easy task. Since the camera field-of-view is relatively small ($5.5^\circ \times 5.5^\circ$), we assume that the illumination inside the sphere was sufficiently homogenous. However, features with a few percent difference in brightness from average have been observed to move when changing the FC pointing, as if associated with the integrating sphere (G.~Kov\'acs, personal communication). Candidates for such features are large scale brightness variations, especially well visible in the FC1 F2 flat field. The experimental setup is a possible source of stray light. During the flat field acquisitions the camera was inside a thermal vacuum chamber, looking out through a window. Reflections off this window may have contributed to the flat field images, though none have been identified with certainty so far. The last two items in our list are hardest to quantify. As the camera design aimed to minimize out-of-field stray light, we assume that its contribution to the flat fields is small. The largest contributor to the non-flat appearance of the lab flat fields is probably in-field stray light. Examples of artifacts we attribute to in-field stray light are the square-shaped brightness enhancement in the center of the F4 and F6 fields, and the more roundish central enhancement in the F7 and F8 fields. To extend the calibration to images of extended sources we need to characterize this form of stray light and to correct the laboratory flat fields for its effects and that of the other factors in the list. We accomplish these tasks in the following sections.

\subsection{Stray light patterns}
\label{sec:stray_light}

\begin{figure}
\centering
\includegraphics[width=11cm,angle=0]{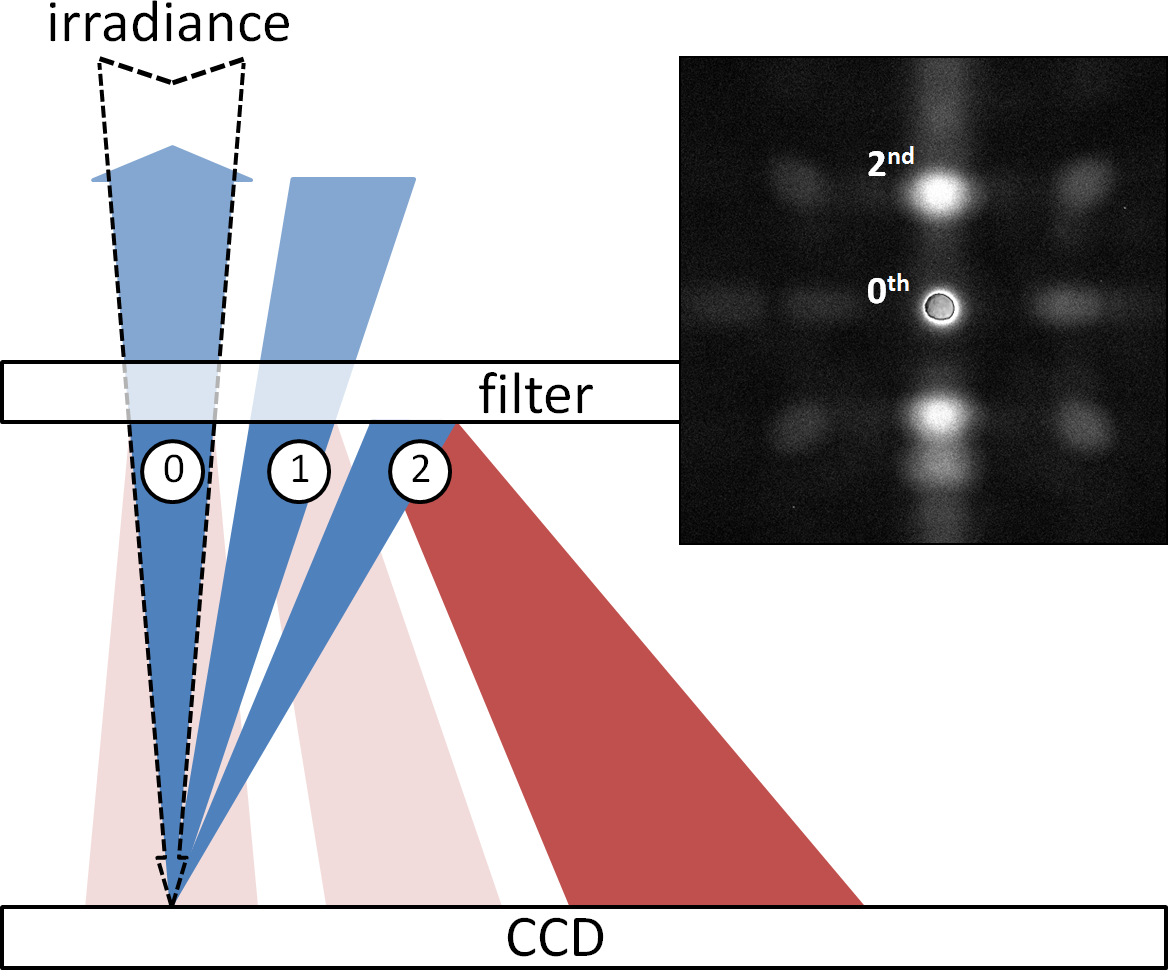}
\caption{Mechanisms of in-field stray light in Framing Camera narrow-band images. The CCD diffracts a fraction of the incoming light (irradiance; dashed line) back towards the filter in different orders (blue), that are partly or fully reflected back towards the CCD (red) depending on incidence angle on the filter. The order of the diffracted light is indicated by a number. See text for details. The inset shows an F6 image of Vesta taken in the {\it Approach} phase, with the area around Vesta (center) enhanced in contrast.}
\label{fig:stray_light_explained}
\end{figure}

It is our task to find the stray light image $\mathbf{I}^i$ to be used in the calibration pipeline for narrow-band (``color'') filter $i$. We write Eq.~\ref{eq:cal_pipeline} as:
\begin{equation}
\mathbf{L}^i = \frac{\mathbf{P}^i - \mathbf{I}^i}{\mathcal{R}^i \mathbf{N}^i},
\label{eq:mod_pipeline}
\end{equation}
where we have defined ``pre-cleaned'' image $\mathbf{P}^i = (\mathbf{A}^i - \mathbf{S} - b)/t_{\rm exp} - \mathbf{D}$. First, we take a closer look at how stray light is generated. Figure~\ref{fig:stray_light_explained} shows the situation for light with a wavelength inside the filter band-pass, which is the narrow range of wavelengths in which the filter is (partially) transparent. The filters are of the interference-type. It is important to realise two things: the filter still reflect 5-15\% of the incident radiation inside the band-pass, and the band-pass of this type of filter shifts with the incidence angle of light. The magnitude of both effects is filter-specific. The CCD is irradiated with light that arrives in a $f/7.5$ convergent beam after passing through the filter. In addition to having a comparatively highly reflective surface, the CCD also acts as a grating due to the presence of gate structures. As such, light is diffracted towards the filter as divergent beams of different orders (indicated in blue in the figure). Subsequently, light is reflected back from the filter towards the CCD (indicated in red). As said, the filters are not completely transparent within the band-pass, so a small fraction of the zeroth order is reflected back, enhancing the signal and enlarging the point spread function. We refer to this light as {\it direct-reflection} stray light. In our example, the same fraction of the first order component is reflected back to the CCD, the remainder passing through the filter. The second and higher orders of reflected light, however, have an incidence angle large enough to shift the filter band-pass completely outside the original wavelength range, making the filter opaque. Hence, essentially all of the incoming light is reflected back towards the CCD. We refer to this light as {\it interference-type} stray light. This stray light is visible as a pattern of spots surrounding Vesta in images taken on approach, when the asteroid was only $\sim$150 pixels large (Fig.~\ref{fig:stray_light_explained}, inset). The spots represent the different orders of diffracted light in both dimensions of the CCD. Direct-reflection-type stray light is not easily recognized or quantified, as it merely increases the intensity on top of the image itself, albeit in a fuzzy fashion. The relative strength of the two types of stray light is not known exactly.

\begin{figure}
\centering
\includegraphics[width=9cm,angle=0]{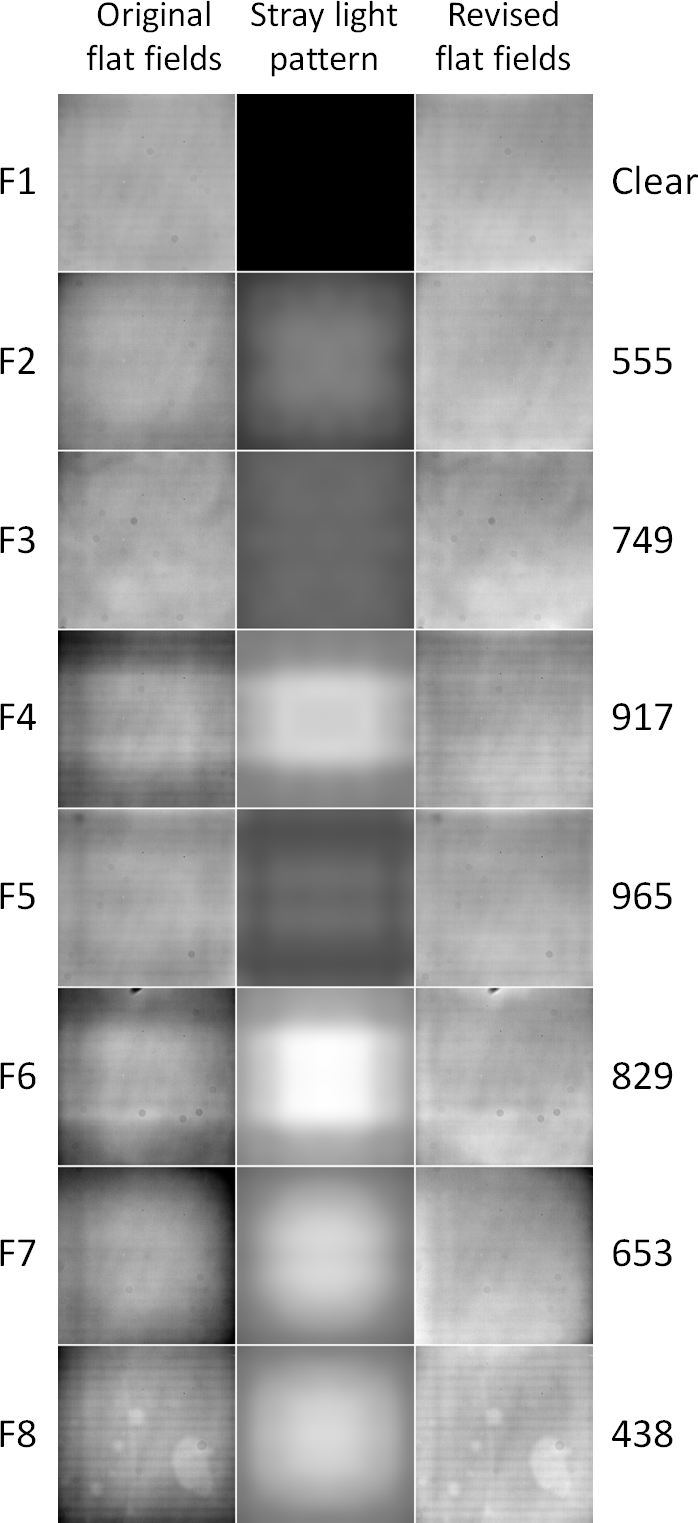}
\caption{Improving the FC2 flat fields. {\bf Left}: Original lab flat fields from Fig.~\ref{fig:FC_flats}. {\bf Center}: Stray light patterns, with black being no stray light (as for F1), and white being the maximum amount of stray light (in F6). {\bf Right}: Revised flat fields (see text). Note that there are actually two versions of the revised F3 flat field, one for each filter wheel rotation direction. Flat fields are displayed as in Fig.~\ref{fig:FC_flats}.}
\label{fig:FC2_flats}
\end{figure}

The in-field stray light pattern is a function of the scene observed, and thus of the clean image. However, we do not know the clean image, so an iterative approach is needed to derive the stray light pattern. We set out to construct a ``typical'' pattern for each filter that is valid for a uniformly bright surface, with the aim to subtract it from the image as observed. For this purpose we combine the lab flat fields of all three FC models shown in Fig.~\ref{fig:FC_flats}. The flat fields show strong similarities for each filter, indicating that stray light acts in the same way for each model. Our patterns are constructed as the median of 12 images, being four images for each of the three camera models. These four images are the original lab flat field in (1) its unaltered form, (2) rotated by $180^\circ$, (3) mirrored, and (4) rotated by $180^\circ$ and mirrored. This approach assumes the stray light pattern has horizontal and vertical symmetry, but allows for diagonal asymmetry. The median pattern is Fourier-filtered to preserve only low frequency variations. The result is free from pixel-to-pixel variations and smooth. Some patterns show relatively strong darkening towards the corners. As in-field stray light preferentially concentrates in the image center, some corner darkening is to be expected. However, at least for some filters the darkening is suspected to be more due to vignetting rather than stray light accumulation in the center (Sec.~\ref{sec:lab_flats}). By reducing the corner darkening of the stray light patterns we can preserve vignetting in the flat fields. We tested reductions of 25\%, 50\%, and 75\%, achieved with help of a polynomial surface fitted to the pattern corners, and determined that 50\% led to the strongest reduction of mosaic seams. The resulting stray light patterns ($\mathbf{I}^i_0$) are shown in Fig.~\ref{fig:FC2_flats} (center). Being the median of normalized flat fields, the stray light patterns are normalized in the center, and this is how they are archived in the Dawn data archive. The figure shows the patterns at their correct relative intensity, that is, the patterns associated with filters that suffer most from stray light are brightest. As such, filter F6 suffers most from stray light, and F3 and F5 least. Some patterns are distinctly squarish in shape (F4, F5, F6), whereas others (F7, F8) are more roundish. The square shape is typical of the interference-type stray light, caused by the second order reflections falling outside the CCD active area. The filters for which the shape is roundish are mostly affected by direct-reflection stray light. Due to our method of generating these patterns, some show distinct features, such as the bands in the F5 pattern. We cannot guarantee that these features are indeed due to stray light, as they may be artifacts intrinsic to the original flat fields. However, they are small in amplitude, and we consider them to be second order effects in our first order correction.

We construct the stray light image $\mathbf{I}^i$ for pre-cleaned image $\mathbf{P}^i$ (Eq.~\ref{eq:mod_pipeline}) from the stray light pattern $\mathbf{I}^i_0$ as:
\begin{equation}
\mathbf{I}^i = p_{\rm C}^i [\mathbf{I}^i_0 - (1 - f^i)],
\label{eq:stray_light}
\end{equation}
in which $p_{\rm C}^i$ is the average charge rate in a square area in the center of $\mathbf{P}^i$, and $f^i$ the fraction of stray light. Before we explain how we retrieve these two parameters we need to resolve an ambiguity. The patterns are smooth, whereas the stray light image $\mathbf{I}^i$ should be affected by pixel-to-pixel sensitivity variations and dust particles on the CCD. In principle, we could multiply the smooth stray light image with a flat field to introduce such variations. However, that would also introduce other variations like those due to vignetting and dust particles on the filters, that should not affect $\mathbf{I}^i$. As it is impossible to separate these effects from the flat fields, we decided to construct smooth stray light images.

\begin{figure}
\centering
\includegraphics[width=11cm,angle=0]{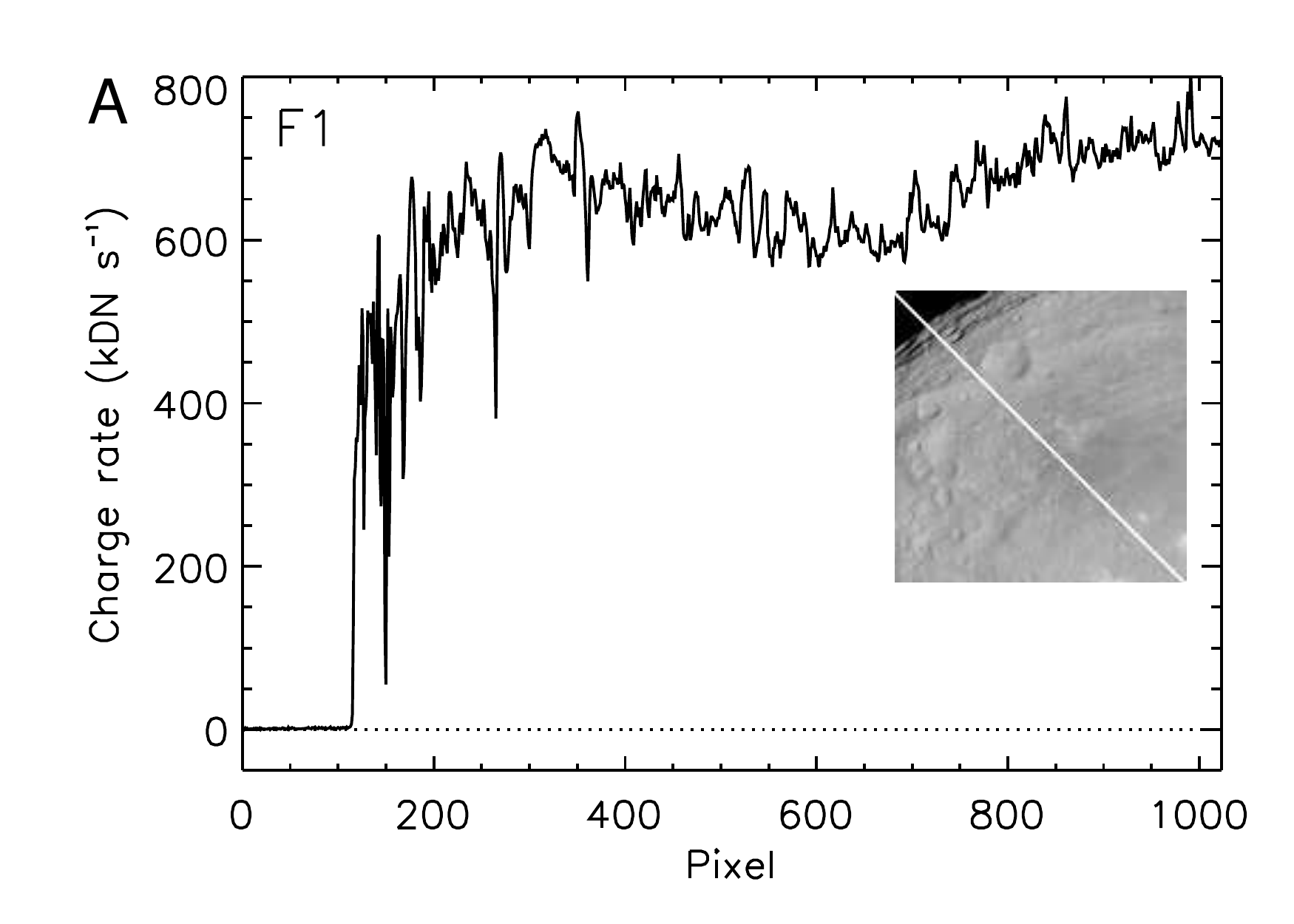}
\includegraphics[width=11cm,angle=0]{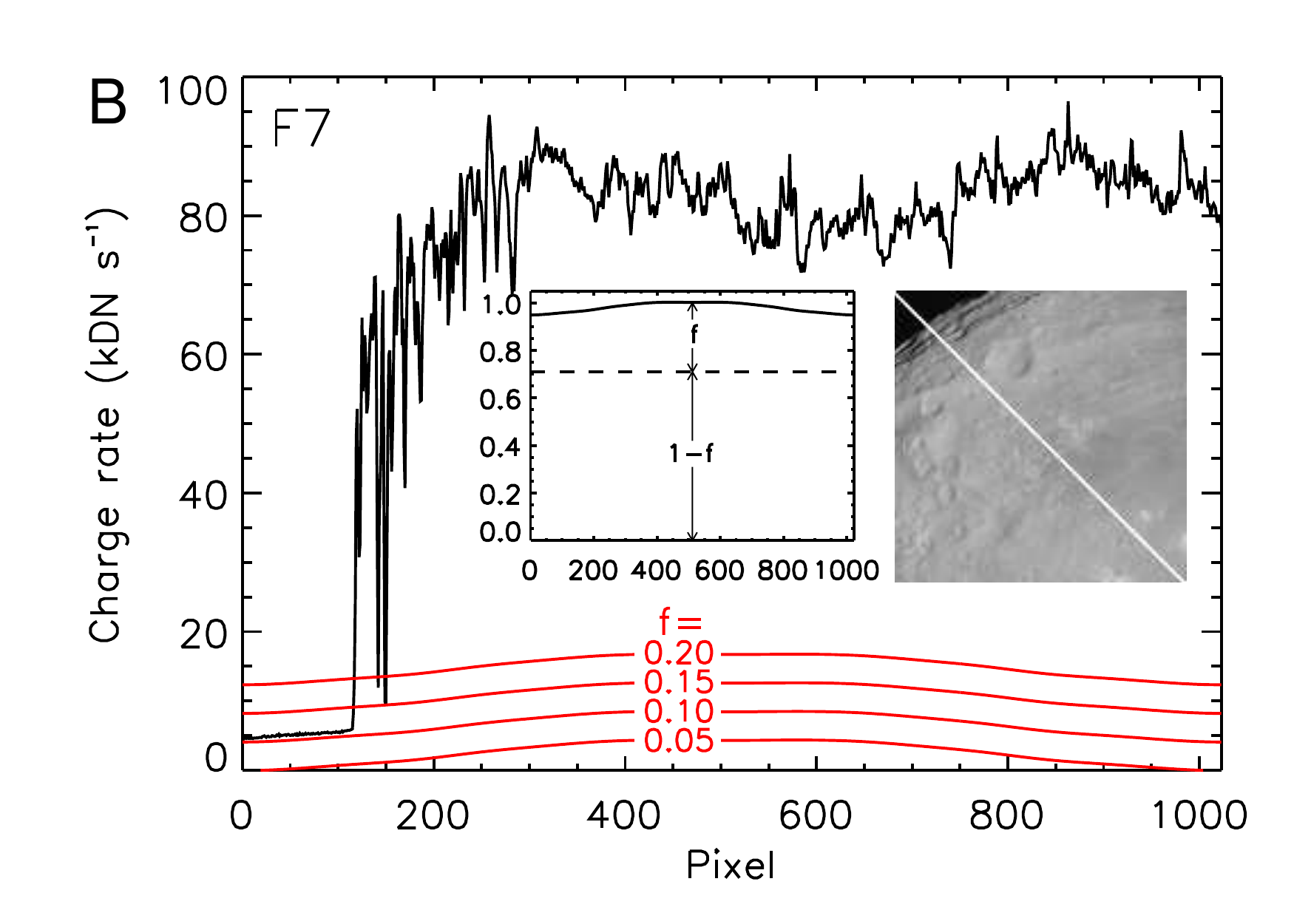}
\caption{Estimating the stray light contribution in pre-cleaned {\it Survey C1} images ($\mathbf{P}^i$ in Eq.~\ref{eq:mod_pipeline}) that have empty space in one corner. The black profile represents the white diagonal in the image inset. {\bf A}. Clear filter image {\bf 4053}. At zero charge rate, the empty corner is unaffected by stray light. {\bf B}. F7 image {\bf 4059}. The red curves represent the stray light contribution for different fractions $f$, calculated with the pattern in Fig.~\ref{fig:FC2_flats} (center column, F7). By comparing these to the profile in the empty image corner we estimate the actual stray light fraction as $f^7 = 0.11$. The plot inset shows the same diagonal of the (normalized) stray light pattern of this filter. $f$ represents the fraction of stray light, where $(1-f)$ is the clean image level.}
\label{fig:stray_light_estimate}
\end{figure}

Our method to quantify the stray light contribution $f^i$ is explained in Fig.~\ref{fig:stray_light_estimate}. We identified images of Vesta that are approximately uniform in brightness, yet show a corner of empty space beyond the limb. The idea is to use any signal in this corner to estimate the stray light contributions to the narrow-band images. We verified that the empty corner in the clear filter image is indeed devoid of signal (Fig.~\ref{fig:stray_light_estimate}A), as this filter is not affected appreciably by stray light. Then we constructed the pre-cleaned image $\mathbf{P}^i$ (Eq.~\ref{eq:mod_pipeline}). We calculated the average charge rate $p_{\rm C}^i$ of pixels (323:700,323:700), which make up a square area in the center of the image that represents a plateau of an almost constant level of stray light (unity) in the normalized pattern $\mathbf{I}^i_0$ of each filter. Of the pre-cleaned charge rate $p_{\rm C}^i$, a fraction $p_{\rm C}^i f^i$ is stray light, while $p_{\rm C}^i (1 - f^i)$ is the true, clean charge rate. The example for filter F7 in Fig.~\ref{fig:stray_light_estimate}B illustrates how we found $f^7$ from the signal in the empty corner of {\it Survey C1} image {\bf 4059}. For several different $f^7$-values we constructed stray light images $\mathbf{I}^7$ through Eq.~\ref{eq:stray_light}, and compared the diagonal profile in the corner to the observations. The best matching profile gives us the true $f^7$ for that image. We repeated this procedure for one diagonal of the images in color cube\footnote{The term ``color cube'' refers to a set of eight images taken through filters F1-F8 or F8-F1 in quick succession.} {\bf 3823}-{\bf 3829} in {\it Survey C0} and two diagonals and one row of cube {\bf 4054}-{\bf 4060} in {\it Survey C1} to obtain the values in Table~\ref{tab:stray_light_faction}. The averages were adopted as the best estimates of $f^i$.

\subsection{Revising the flat fields}
\label{sec:flat_revision}

With the results of our stray light investigation we are ready to improve the laboratory flat fields. First we subtracted the stray light from the lab flat fields using the patterns and fractions derived in Sec.~\ref{sec:stray_light}, and used the revised flat fields to calibrate all images of Vesta acquired during the {\it High Altitude Mapping Orbit} ({\it HAMO) C1}. These images (320 clear filter images and 278 images per color filter) were then stray-light subtracted and photometrically corrected using the clear filter model from \citet{S13b}. We averaged all images with an average phase angle $<50^\circ$ for each filter (illuminated pixels only). We excluded large phase angles to avoid shadows. The averaged images were not flat, but showed distinct artifacts due to the factors listed in Sec.~\ref{sec:lab_flats}. To verify that these artifacts are not residual albedo and shadowing features, we confirmed their presence in averages of the {\it HAMO C6} data set. To further improve the flat fields we decided on an iterative approach. In the first iteration we fitted polynomial surfaces to the averages, which we then multiplied with the flat fields to yield a new generation of flat fields. As described in the next section, we derived new photometric models for each of the color filters using the {\it C1} data set. In the second iteration we re-calibrated all color images using the newly revised flat fields and applied the filter-specific photometric corrections. Again, we fitted a polynomial surface to the averages, and constructed the final revision of the flat fields, shown in Fig.~\ref{fig:FC2_flats} (right). These flat fields are identified with $\mathbf{N}^i$ in Eq.~\ref{eq:cal_pipeline}. Note that they are normalized to an area in the center with coordinates (579:619, 493:553), which is the spot covered by the monochromator beam during the lab calibration \citep{Si11}.

The photometrically corrected averages also served a different purpose. Close inspection revealed the presence of bad image pixels and shifted dust specks on the filters. Bad pixels are not responsive to light and are best replaced with the average of surrounding good pixels. Their locations within the image frame are recorded in a file uploaded to the Dawn data archive. Concerning the dust specks, these are out of focus and can clearly be identified in the flat fields, typically as circles of 40 pixels in diameter. We identified a dust speckle in filter F8 that had moved by about 7 pixels since launch. In addition, we found that the position of the dust speckles in F3 depends on the direction of the filter wheel rotation (clockwise or anti-clockwise). That is, if an F3 image acquisition follows that an F2 image, the specks are shifted with respect to their position in an F3 image following an F4 image. Thus we prepared two flat fields for filter F3, one for each filter wheel rotation direction.

\section{Testing the calibration}

We evaluate the results our stray light correction and flat field improvements in two different ways. First, we look at the quality of the absolute calibration by comparing an FC2-derived spectrum of Vesta with earlier observations. Second, we judge the quality of color ratio mosaics, which are strongly sensitive to stray light and inaccuracies in the flat field. To this end we projected images with the USGS Integrated Software for Imagers and Spectrometers ISIS3 \citep{B12}, using the \citet{G11} shape model (post-{\it LAMO} version). Prior to projection, small pointing errors in the SPICE kernels were corrected to achieve an optimal match to the shape model.

\subsection{Radiometric calibration}
\label{sec:rad_cal}

\begin{figure}
\centering
\includegraphics[width=10cm,angle=0]{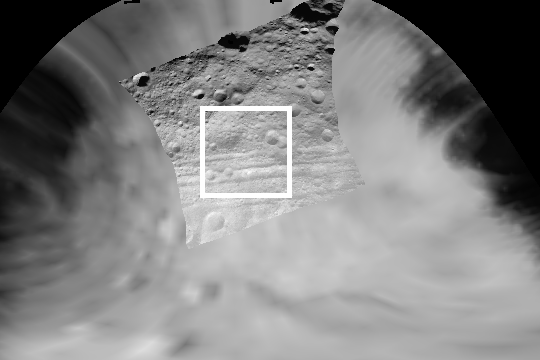}
\includegraphics[width=11cm,angle=0]{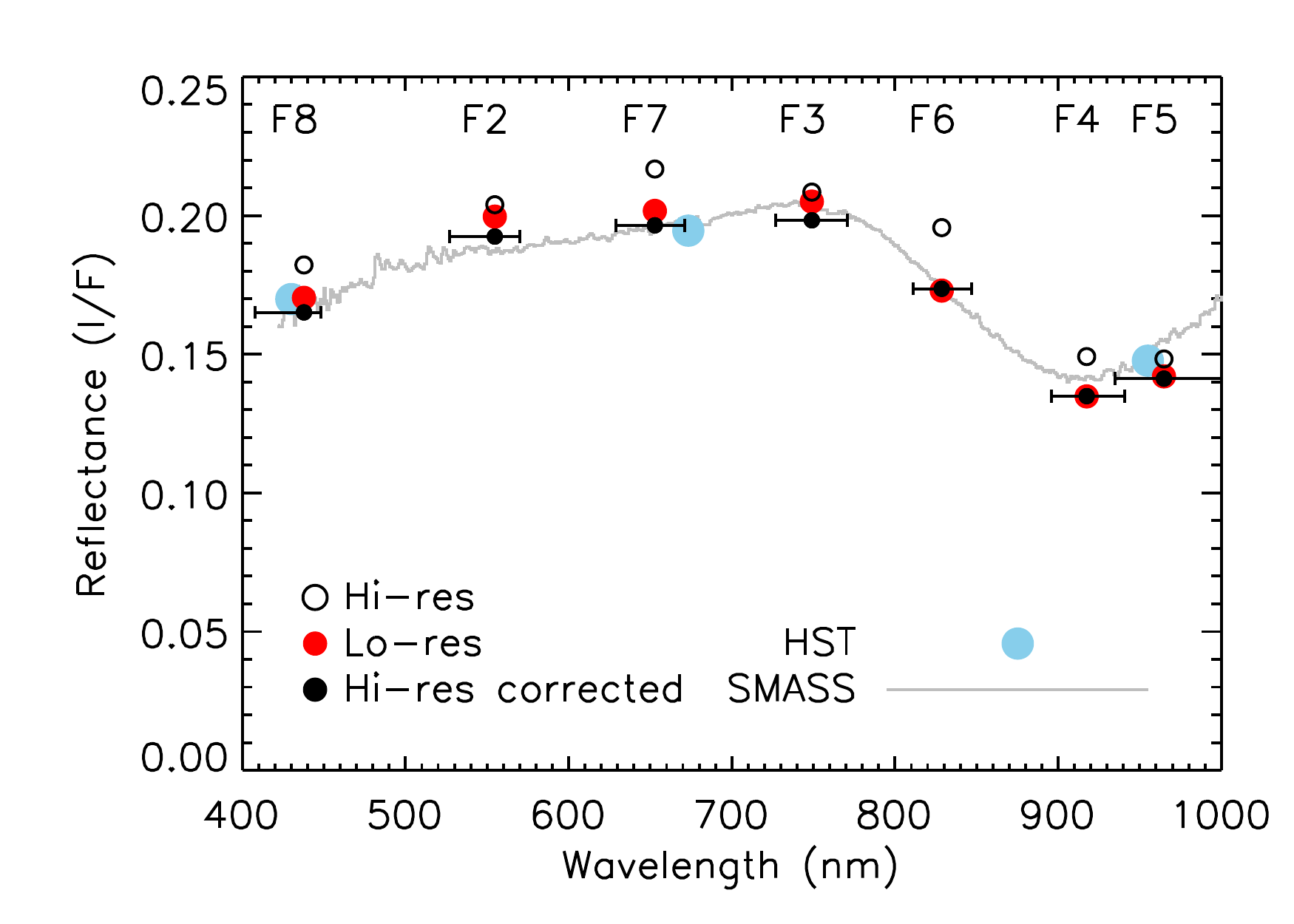}
\caption{Comparison of the Vesta surface reflectance in low- and high-resolution images acquired at approximately the same phase angle during the {\it RC1} and {\it C0} campaigns, respectively. Shown is the area with latitude $(-70^\circ,50^\circ$) and longitude $(-30^\circ,150^\circ$), with a hi-res image projected on top of a lo-res image (both F2). The plot compares the average reflectance in the box outlined in white in the projected lo- and hi-res images, the latter both before and after subtraction of in-field stray light. All images were calibrated with the revised flat fields. The SMASS \citep{X95} and HST \citep{L10} data were scaled to match the corrected hi-res reflectance. The horizontal error bars denote the FWHM of the filter transmission curves (only shown for one data set for clarity).}
\label{fig:comparison_RC1_C0}
\end{figure}

The first opportunity to observe Vesta through the FC narrow band filters came with the first rotational characterization campaign {\it RC1} on 30 June 2011. During {\it RC1} the Vesta disk was small enough (50-60 pixels across) for the (tiny) image of Vesta itself to be free from interference-type stray light (see the example in Fig.~\ref{fig:stray_light_explained}). When we compare the surface reflectance reconstructed from low-resolution {\it RC1} images with that from higher resolution images acquired in orbit, we can quantify the stray light contribution and evaluate the quality of the absolute calibration described by \citet{S13a}. We selected an area on the surface that was observed at an almost identical phase angle at the end of the {\it Approach} phase, in pre-{\it Survey} orbit {\it C0}. The images selected for the analysis are {\it RC1}: {\bf 2341}-{\bf 2347} and {\it C0}: {\bf 3768}-{\bf 3774}, with average center-of-image phase angles of $26.2^\circ$ and $26.7^\circ$, respectively. We calibrated the images to reflectance with the revised flat fields without stray light subtraction, and we prepared an additional set of calibrated high-resolution images with stray light subtracted. Figure~\ref{fig:comparison_RC1_C0} compares the low-resolution (lo-res) reflectance with the high-resolution (hi-res) reflectance, both before and after stray light subtraction, along with a Vesta spectrum from the Small Main-Belt Asteroid Spectroscopic Survey (SMASS; \citealt{X95}) and Hubble space telescope data (HST; \citealt{L10}). The slope of the visible part of the spectrum is not expected to vary much, neither with phase angle (phase reddening) or over the surface, but the depth of the 1~\textmu m absorption band does vary significantly over the surface \citep{R12a,L13}, so we can expect discrepancies for filters F4 and F5 when comparing the reconstructed reflectance with the aforementioned data.

Comparing the reflectances in Fig.~\ref{fig:comparison_RC1_C0} we infer that the four filters in the visible wavelength range (F8, F2, F7, F3) suffer from both the interference and direct types of stray light, since the corrected hi-res reflectance (black bullets) is slightly below the lo-res reflectance (red bullets). On the other hand, the three near-IR filters (F6, F4, F5) are mainly affected by interference-type stray light and not direct-reflection stray light, as the lo-res and corrected hi-res reflectances are identical. The difference between the lo-res and (uncorrected) hi-res reflectance is the amount of in-field-stray light in the hi-res images. The agreement with the SMASS and HST data is good, and slightly better for the corrected hi-res images than the lo-res images, an important indication that our correction is successful. The F2 and F3 reflectances are slightly above and below the SMASS spectrum, respectively. It is not clear whether this represents true surface variegation, or errors in the calibration of either FC2 or the SMASS spectrometer. We therefore adopt these discrepancies as the expected uncertainties of the full calibration pipeline of the \citet{S13a} radiometric calibration including the stray light removal method described in this paper. The F4 and F5 reflectances are slightly below the SMASS/HST spectrum, but this almost certainly due to surface variegation. Our reconstructed Vesta spectrum matches the SMASS/HST spectra better than that in \citet{L13}, which we take as evidence that our stray light correction is much more accurate.

Now that we have confidence in our stray light correction and the absolute calibration of Vesta images we need to revisit several aspects of the first paper in this series \citep{S13a}. First, the responsivities were derived on the assumption that stray light does not affect observations of point sources. In light of the results in this paper, we cannot exclude that direct-reflection type stray light does play a role for filters F2, F3, F7, and F8. If stray light indeed increases the point source flux, then the derived responsivities are too high by a few percent, depending on the ratio of direct-reflection to interference-type stray light. The reflectance of stray light-corrected images obtained through these responsivities would then be too low, especially for filters F7 and F8, which have more stray light than F2 and F3. While such a trend is not obvious in Fig.~\ref{fig:comparison_RC1_C0}, we cannot exclude that direct-reflection stray light contributes to observations of point sources. Second, the first paper derived narrow band responsivities for targets with a solar reflectance spectrum. We can fine-tune the responsivities for Vesta using the SMASS spectrum \citep{X95}, arriving at the responsivity factors in Table~\ref{tab:responsivities}. The consequences are negligible for all filters save F5 and F8, whose responsivity profiles are strongly skewed. Applying the revised factors moves the F5 reflectance up by 2\% and the F8 reflectance down by 2\%. Third, we can now develop an expression for the clear filter responsivity factor that includes the effective width, enabling us to calculate the clear filter reflectance for Vesta. This factor, in units of [J$^{-1}$ m$^2$ nm sr], is
\begin{equation}
\mathcal{R}^1 = \frac{A \Omega_{\rm px} c^1 \Delta\lambda^1 \int_{\lambda^1_{\rm lo}}^{\lambda^1_{\rm hi}} r^1(\lambda) F_\odot(\lambda) \mathrm{d}\lambda}{\int_{\lambda^1_{\rm lo}}^{\lambda^1_{\rm hi}} F_\odot(\lambda) \mathrm{d}\lambda},
\label{eq:responsivity}
\end{equation}
in which $A = 3.41\times10^{-4}$~m$^2$ is the FC aperture, $\Omega_{\rm px} = 8.69\times10^{-9}$~sr is the solid angle of a single pixel, and $c^1 = 1.11$ a correction factor that brings the ground calibration in line with the in-flight observations (see first paper for details). For the integration limits $\lambda^1_{\rm lo}$ and $\lambda^1_{\rm hi}$ we adopt 400 and 1100~nm, respectively, these approximately being the limits of the CCD responsivity. The reflectance, expressed as the radiance factor $r_{\rm F}$ (or I/F), for pixel ($x,y$) is
\begin{equation}
(r_{\rm F})_{xy}^i = \pi d^2_{\rm V} L^i_{xy} / F^i_\odot,
\label{eq:reflectance}
\end{equation}
As discussed by \citet{S13a}, it is well defined for narrow-band color filters ($i=2$-8), but not the clear filter ($i=1$). However, we can now predict the average clear filter reflectance in the box in Fig.~\ref{fig:comparison_RC1_C0} as
\begin{equation}
\left<r_{\rm F}^1\right> = \frac{\int_0^\infty r_{\rm F,Vesta}(\lambda) r^1(\lambda) \mathrm{d}\lambda}{\int_0^\infty r^1(\lambda) \mathrm{d}\lambda} = 0.184,
\end{equation}
adopting the (scaled) SMASS spectrum for $r_{\rm F, Vesta}$. Naturally, this value is only valid at this particular phase angle ($\sim26^\circ$). $r^1(\lambda)$ is the clear filter responsivity profile in DN J$^{-1}$ \citep{Si11}. By comparing this value with the actual observed box average in the clear filter (image {\bf 3767}), we retrieve an effective width (FWHM) of $\Delta\lambda^1 = 682$~nm, where we used $F^1_\odot = 1.347$ [W m$^{-2}$ nm$^{-1}$], the effective solar flux at 1~AU in the clear filter. This corresponds to a responsivity factor of $\mathcal{R}^1 = 3.49 \times 10^7$ [J$^{-1}$ m$^2$ nm sr], which is only valid for Vesta. The clear filter effective width and responsivity factor are expected to be different for Ceres.

\subsection{Image mosaics}

\begin{figure}
\centering
\includegraphics[width=\textwidth,angle=0]{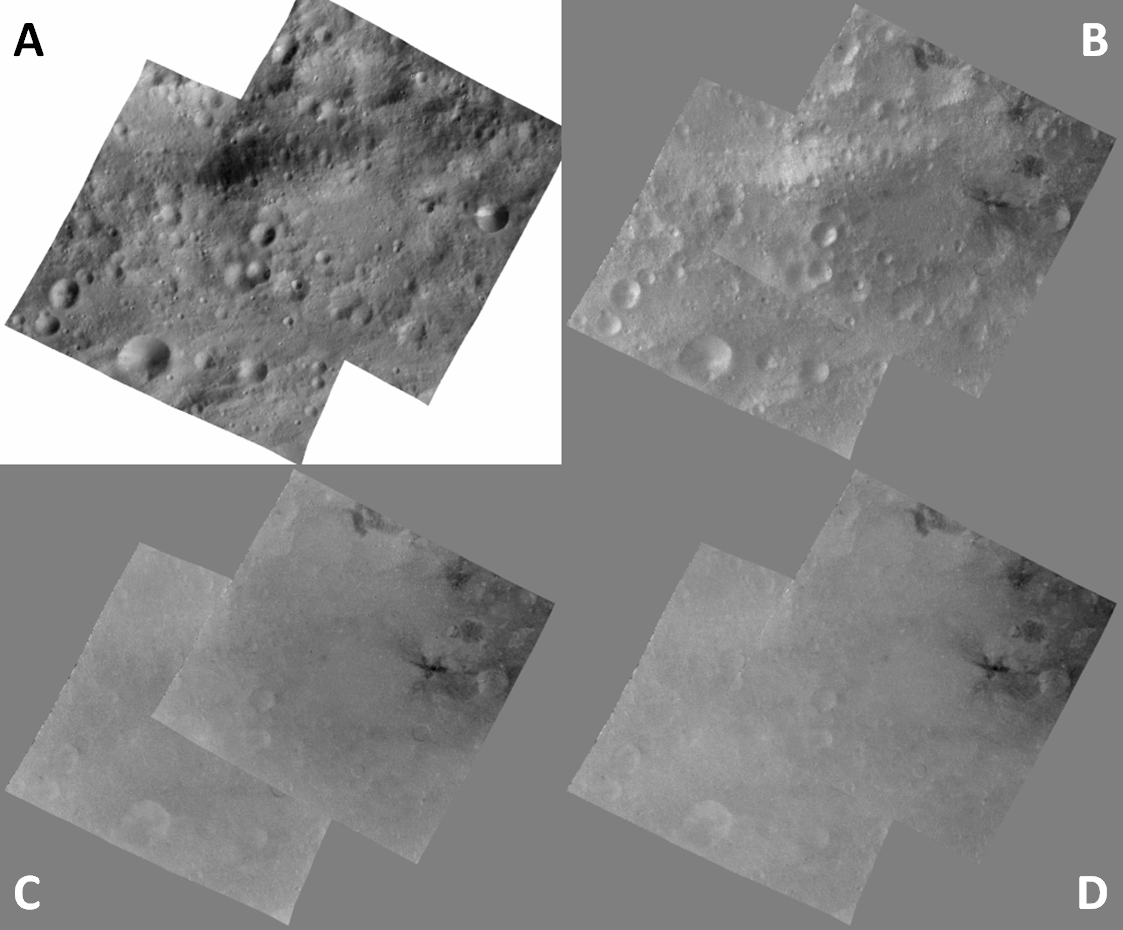}
\caption{The effectiveness of the stray light removal method can be evaluated from two-image mosaics of an area around ($-15^\circ$, $285^\circ$E) (Claudia coordinate system; \citealt{Ru12}), which show the ratio of the reflectance in the blue filter (F8) over that in the clear filter. {\bf A}: Clear filter mosaic, shown for reference. {\bf B}: Ratio mosaic of images calibrated with the lab flat fields and not corrected for stray light. {\bf C}: Ratio mosaic of stray light-subtracted images calibrated with stray light-subtracted lab flat fields. This mosaic reveals defects in the stray light-subtracted flat fields. {\bf D}: Ratio mosaic of stray light-subtracted images calibrated with the revised flat fields. Black and white in the ratio mosaics are scaled to $\pm 10$\% of the median brightness. The reflectance is photometrically corrected with the model in Sec.~\ref{sec:phot_mod}. All mosaics are in (identical) equirectangular projection and have the same image priority.}
\label{fig:mosaic}
\end{figure}

\begin{figure}
\centering
\includegraphics[width=\textwidth,angle=0]{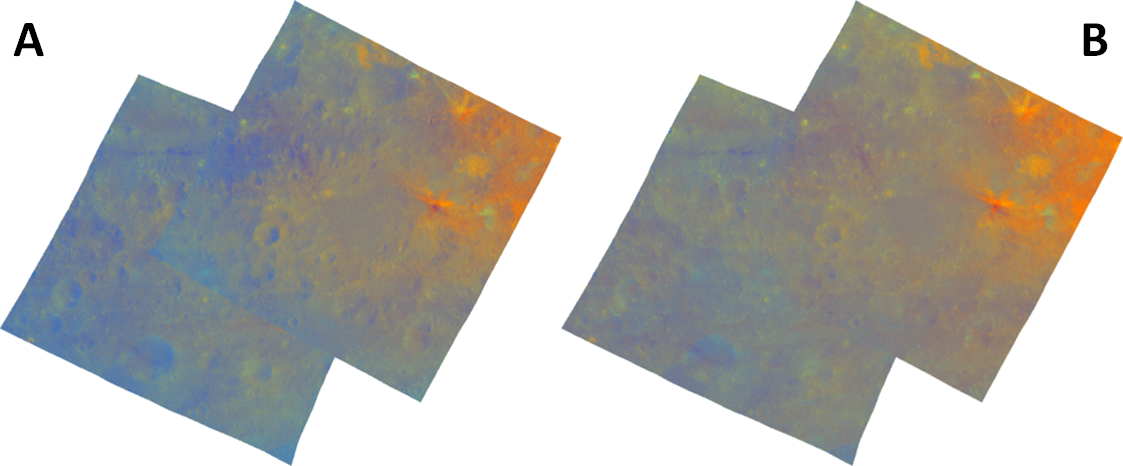}
\caption{The terrain in Fig.~\ref{fig:mosaic} in the Clementine color scheme, in which red is 750~nm / 430~nm, green is 750~nm / 920~nm, and blue is the inverse of red. {\bf A}: Mosaic of images calibrated with the lab flat fields and not corrected for stray light. Topography is associated with an excess of blue. {\bf B}: Mosaic of stray light-subtracted images calibrated with the revised flat fields. Craters are no longer recognizable by their topography, and the color matching between the two images has improved.}
\label{fig:Clem_mosaic}
\end{figure}

We evaluate the effectiveness of the stray light removal method and the flat field revision by visual inspection of an image mosaic. We chose a {\it HAMO C1} color cube at random, and constructed mosaics from images from this cube and the subsequent one. The images in these two cubes are numbered {\bf 7123}-{\bf 7138} (16 images). First, we constructed ratio images from the images in one cube, and then combined the ratio images of both cubes into a mosaic to avoid the appearance of seams. Figure~\ref{fig:mosaic} shows mosaics of the ratio of the photometrically corrected reflectance in the blue filter (F8) to that in the clear filter (F1). The photometric correction employs the photometric models in Table~\ref{tab:phot_mod_coef}. The clear filter has virtually no stray light, so the mosaics show the consequences of stray light in F8. Three color ratio mosaics are shown with the images calibrated in three different ways to visualize the artifacts due to in-field stray light and flat field imperfections. When we calibrate the images with the original flat fields as acquired in the laboratory (Fig.~\ref{fig:FC2_flats}, left), we find that the mosaic is dominated by brightness changes associated with topography. Strong brightness gradients across craters and ridges obscure color variations intrinsic to the surface. These gradients are spurious and due to in-field stray light. This can be understood as follows. A crater that is illuminated at an angle will have one side that is brighter, and an opposite side that is darker than the surroundings. Assuming the crater is small relative to the full image, the stray light contribution will be approximately constant over the crater. However, fraction of stray light relative to the unpolluted signal level will be much larger on the dark side. Therefore, after division with an image that is less affected by stray light (in this case a clear filter image), the dark side of the crater will stand out as bright. At the same time, the bright side will be a little darker than average, exactly as we see in Fig.~\ref{fig:mosaic}. After calibrating the images by subtracting stray light from both the images and the flat fields as described in Sec.~\ref{sec:stray_light}, craters are no longer recognizable, allowing intrinsic color variations to stand out clearly. However, the individual ratio images that make up the mosaic do not match well where they overlap, especially in the corners. This is due to imperfections in the lab flat fields, as explained in Sec.~\ref{sec:flat_revision}. When using the revised flat fields, the mismatch between the individual ratio images is eliminated to a large degree, proving the success of our calibration method. We show the same terrain in Fig.~\ref{fig:Clem_mosaic} in the Clementine color scheme \citep{R12a}, where red is the slope in the visible wavelength range, green is the 1~\textmu m band depth, and blue is the inverse of red. Also here, the improvement resulting from using the revised flat fields is obvious. Note how craters can be recognized in the uncorrected mosaic on the left, their dark side being blue.

\section{Photometric modeling}
\label{sec:phot_mod}

\subsection{Model}

The images calibrated according to \citet{S13a} in combination with the method outlined in this paper are radiometrically accurate to a high degree. Here we develop a photometric model for the reflectance in each filter for the purpose of photometric image correction. Photometrically corrected images can be combined into a seamless mosaic. We consider a simple photometric model for the reflectance $r_{\rm F}$ (radiance factor or I/F; \citealt{H81}) that can be separated in a phase function and a disk function:
\begin{equation}
r_{\rm F} = A_{\rm eq}(\alpha) D(\mu_0, \mu, \alpha),
\label{eq:eq_albedo}
\end{equation}
where $A_{\rm eq}$ is the equigonal albedo, $\alpha$ is the phase angle, and $D$ is the disk function \citep{S11}. For convenience we have defined $\mu_0 = \cos\iota$ and $\mu = \cos\epsilon$, with $\iota$ and $\epsilon$ the incidence and emergence angles, respectively. The disk function essentially describes the brightness variations over the planetary disk. We adopt the Akimov model:
\begin{equation}
D({\alpha, \beta, \gamma}) = \cos \frac{\alpha}{2} \cos \left[ \frac{\pi}{\pi - \alpha} \left( \gamma - \frac{\alpha}{2} \right) \right] \frac{(\cos \beta)^{\alpha / (\pi - \alpha)}}{\cos \gamma}.
\label{eq:Akimov}
\end{equation}
It employs the photometric latitude $\beta$ and longitude $\gamma$ that depend on the incidence, reflection, and phase angles as follows:
\begin{equation}
\begin{split}
\mu_0 & = \cos \beta \cos (\alpha - \gamma) \\
\mu & = \cos \beta \cos \gamma
\end{split}
\end{equation}
This parameter-free version of the Akimov function was derived theoretically for an extremely rough surface that is slightly randomly undulated \citep{S03,S11}. The Akimov disk model has no parameters and was found to work well for the surface of Vesta \citep{S13a}.

\begin{figure}
\centering
\includegraphics[width=5cm,angle=0]{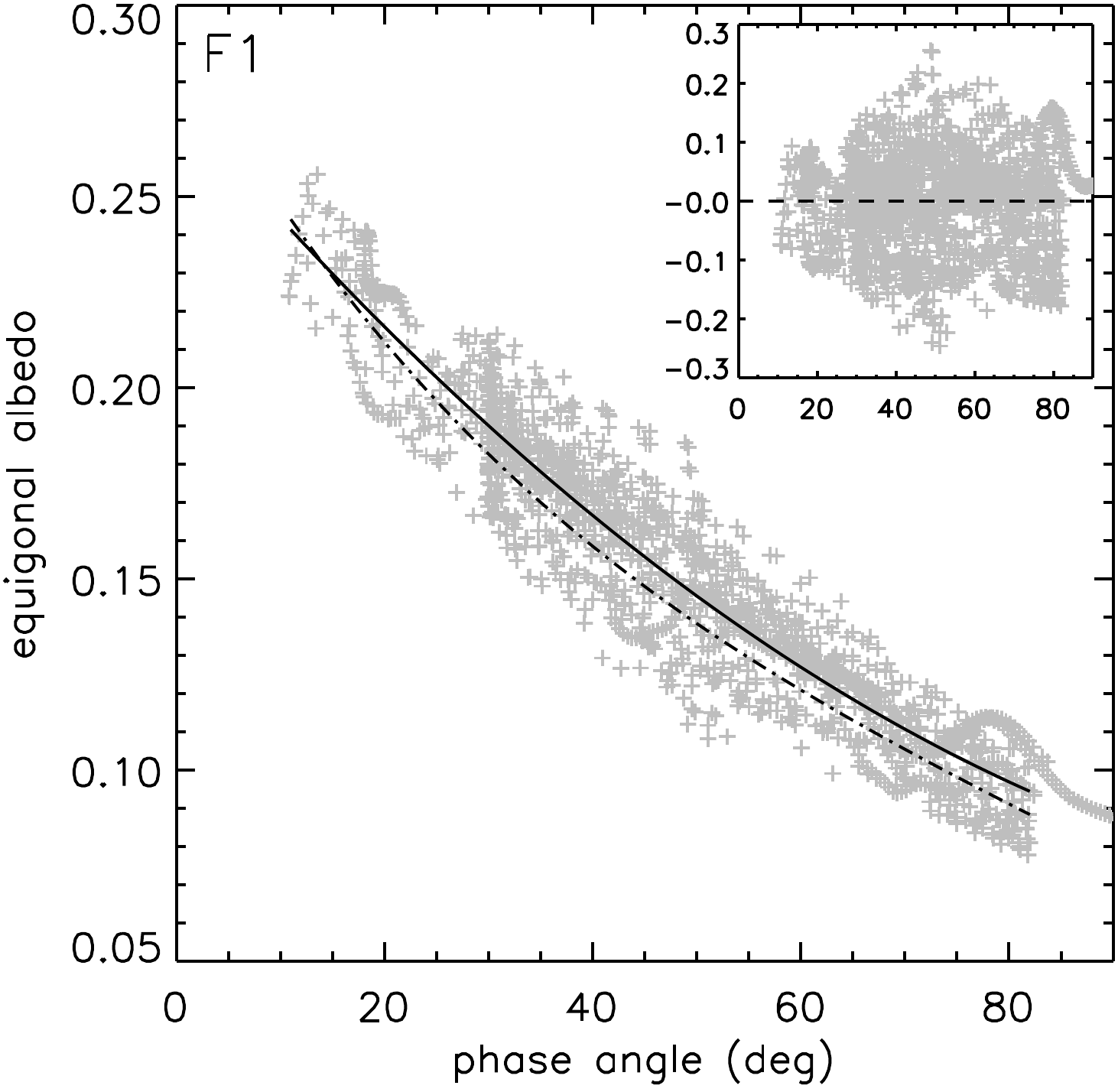}
\includegraphics[width=5cm,angle=0]{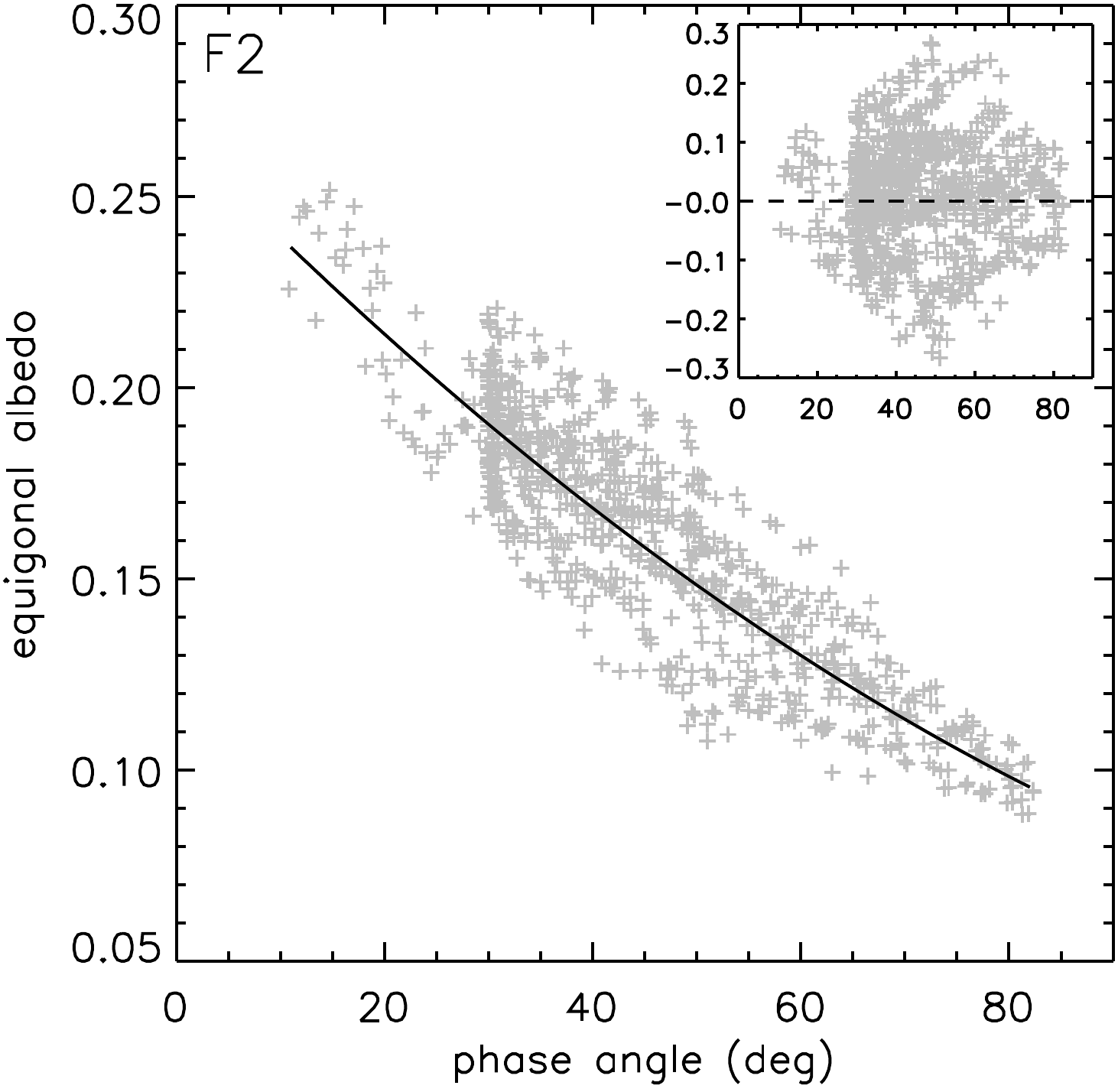}
\includegraphics[width=5cm,angle=0]{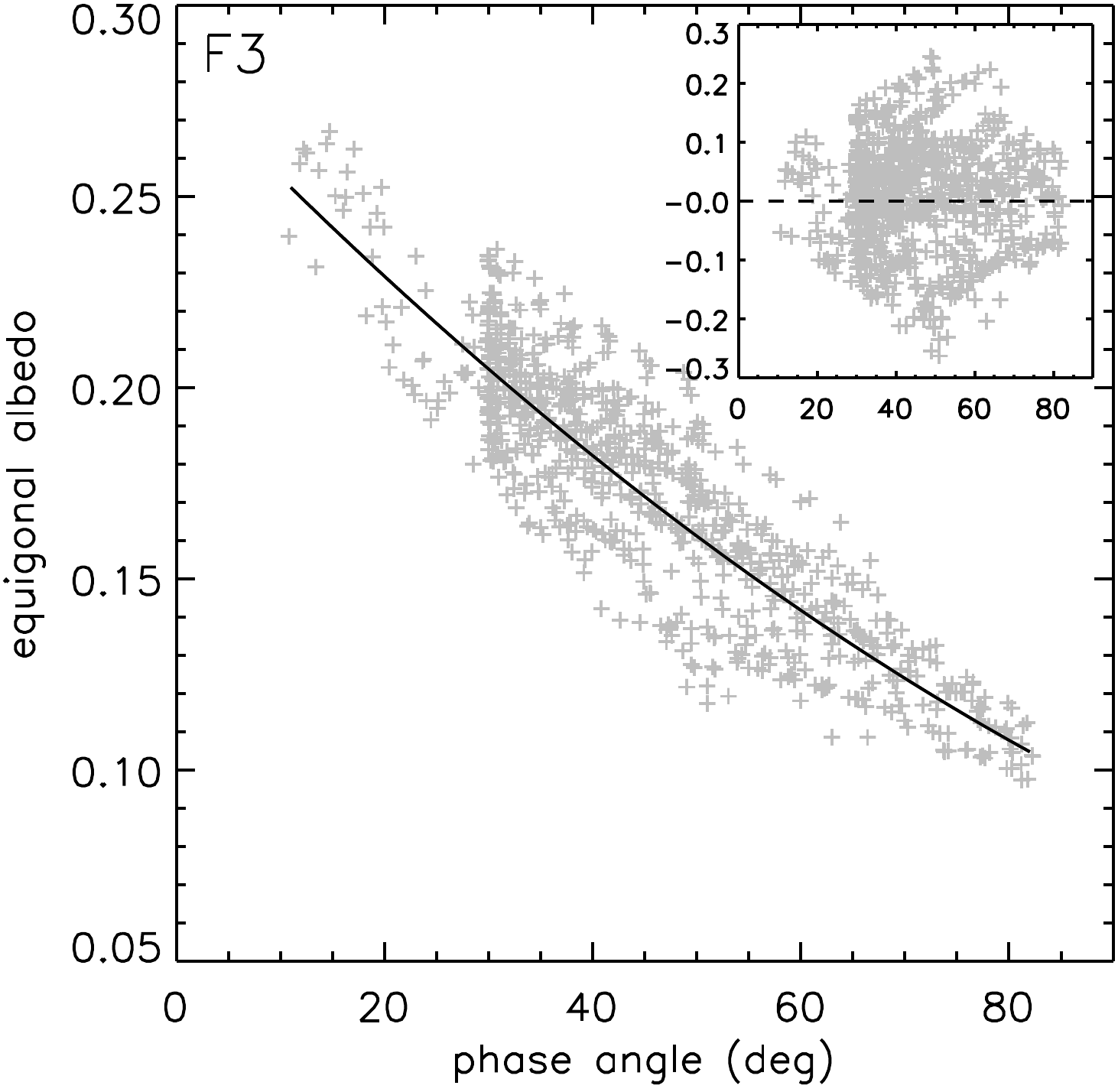}
\includegraphics[width=5cm,angle=0]{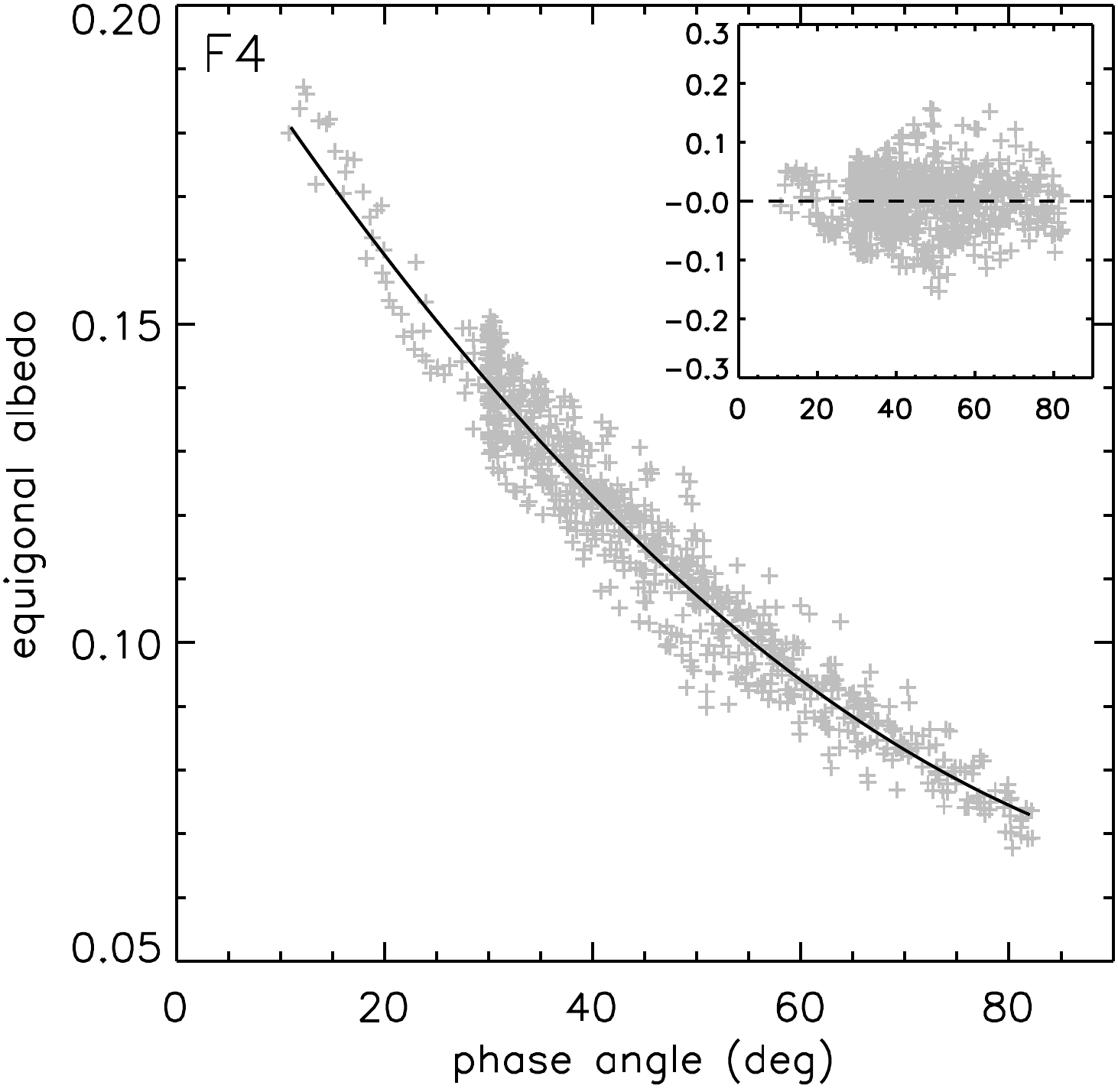}
\includegraphics[width=5cm,angle=0]{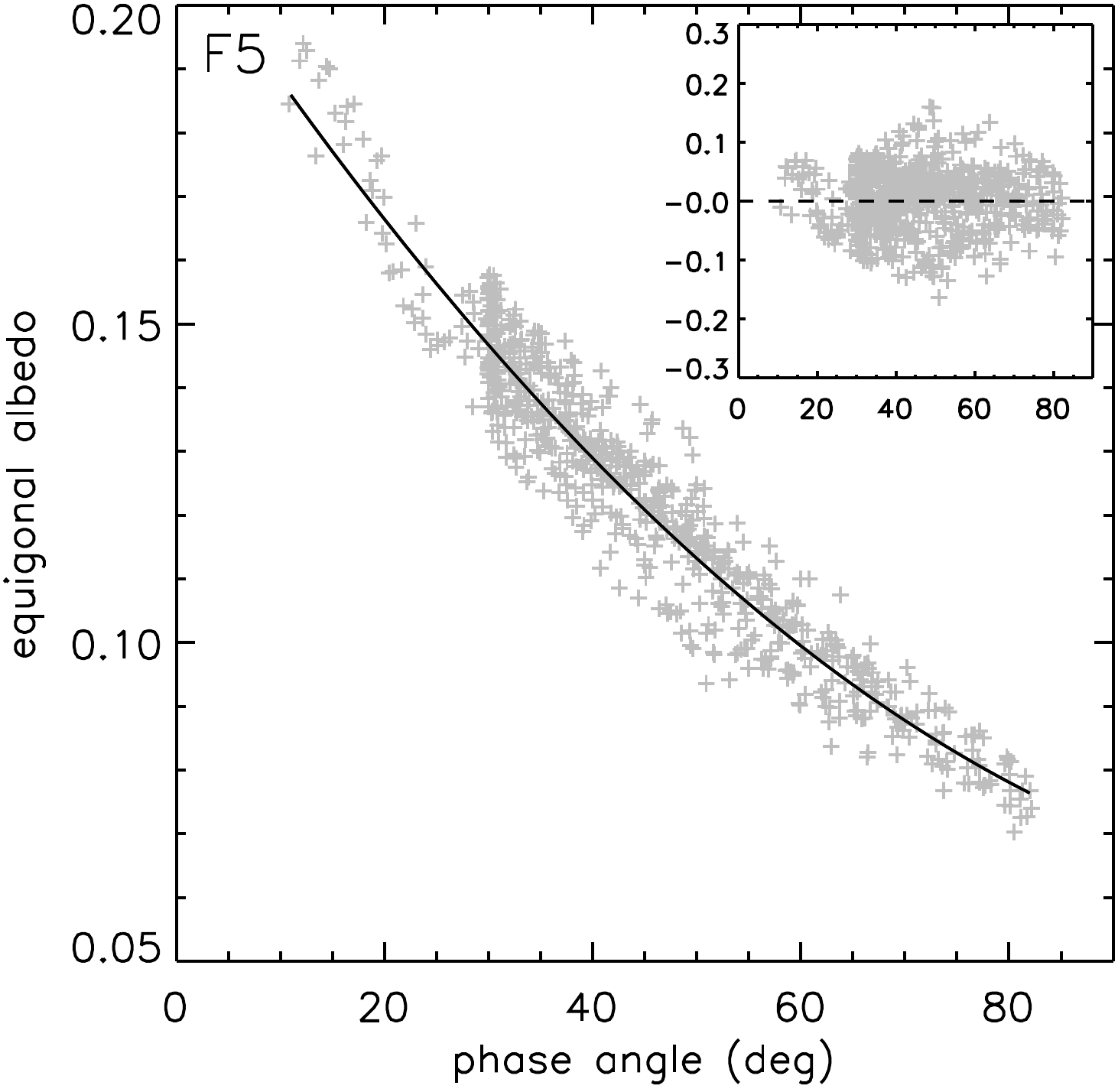}
\includegraphics[width=5cm,angle=0]{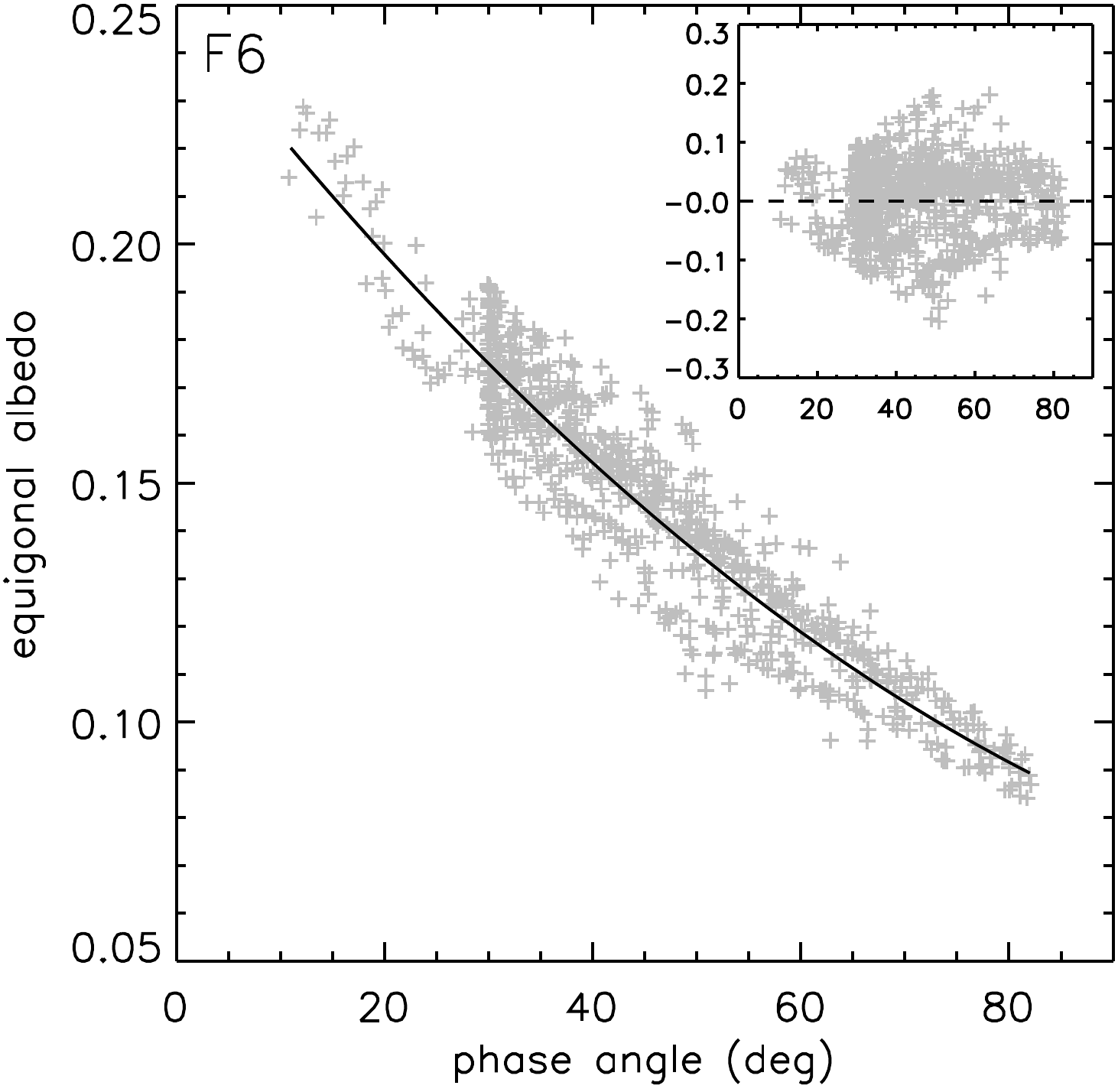}
\includegraphics[width=5cm,angle=0]{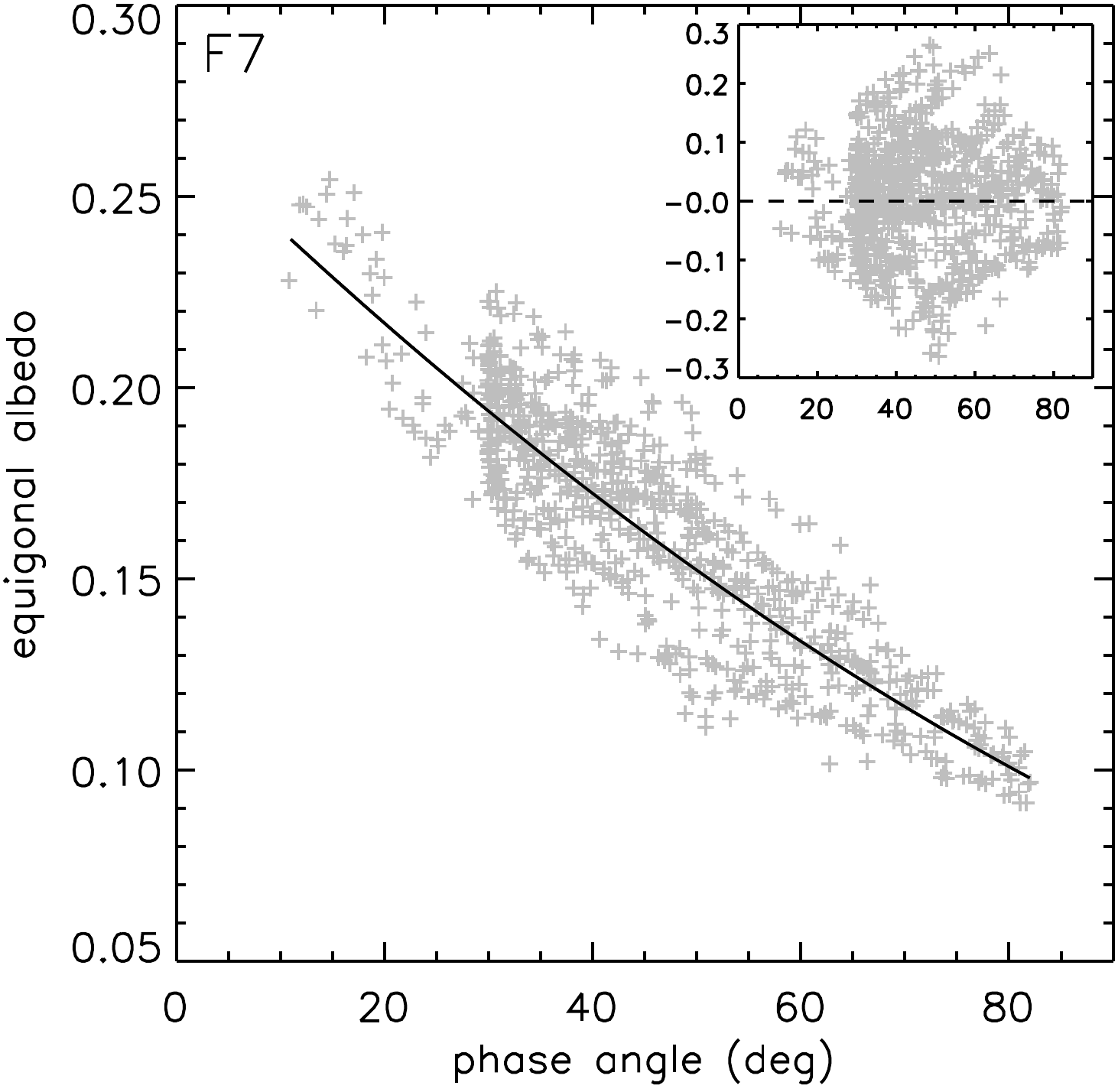}
\includegraphics[width=5cm,angle=0]{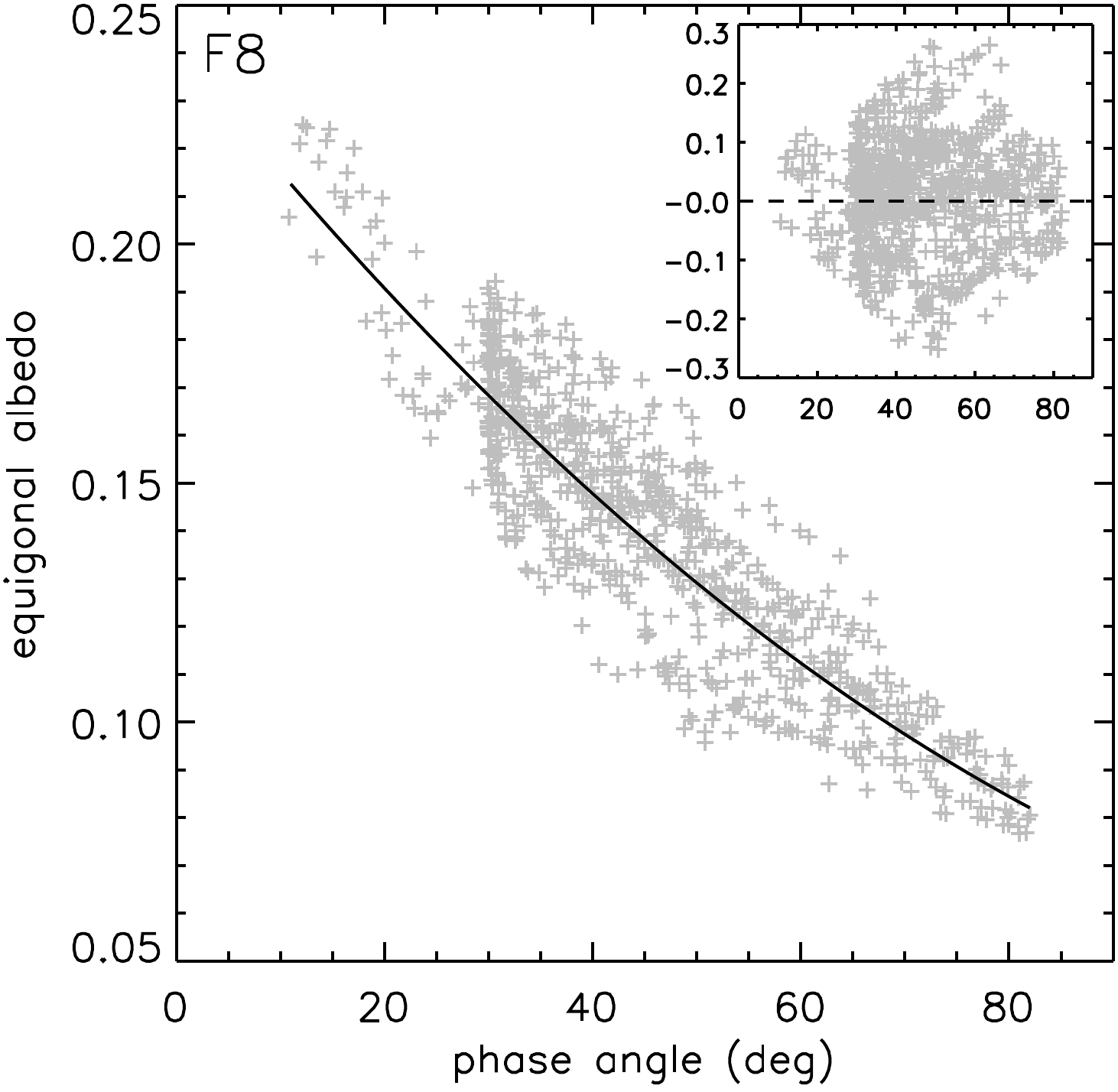}
\caption{Phase curves of the equigonal albedo in each of the filters for all {\it Survey} and {\it HAMO C1/C6} images. Each symbol (+) is the median albedo of an image photometrically corrected with the parameterless Akimov function. The drawn lines are best-fit parabolas, with the coefficients in Table~\ref{tab:phot_mod_coef}. The inset shows the relative residuals. The dash-dotted line in the F1 plot is the \citet{S13b} $4^{\rm th}$-order polynomial photometric model.}
\label{fig:phase_curves}
\end{figure}

We derive a phase function for each filter by analyzing all images acquired during the {\it Survey} and the {\it HAMO C1} and {\it C6} campaigns. The data suffer from observation bias in two ways: In {\it Survey}, the northern latitudes were not yet visible, and during {\it HAMO} the (bright) south pole was not observed at high phase angles. Each image was calibrated with the revised flat fields, stray light subtracted, and photometrically corrected with the Akimov disk function, with the photometric angles calculated with ISIS3 \citep{B12} from a shape model by \citet{G11}. We used the post-{\it LAMO} version of this model, and verified that its resolving power is approximately half that of the {\it HAMO} images (the smallest details in the shape model are twice the size of those in the images). The result of the correction is the equigonal albedo. For each image we calculated the average equigonal albedo for pixels with $r_{\rm F} > 0.01$ and $(\iota, \epsilon) <80^\circ$. We adopt a simple expression to model the albedo averages, a parabola, which nevertheless fits the data well over the full phase angle range of our data set. The resulting phase curves for all filters are shown in Fig.~\ref{fig:phase_curves}, with the fit parameters listed in Table~\ref{tab:phot_mod_coef}. Inspection of the residuals reveals that photometric variability is smallest in the two near-IR filters F4 and F5. The depth of the 1~\textmu m band is known to be strongly variable over the surface of Vesta \citep{R12a}. Hence, it is mainly the visible reflectance that varies (F8, F2, F7, F3), rather than the reflectance inside the band (F4, F5), not only in absolute, but also in relative terms. The F4 phase curve is steepest of all curves, which is consistent with the fact that the reflectance in this filter is lowest (cf.\ Fig.~\ref{fig:comparison_RC1_C0}).

\begin{figure}
\centering
\includegraphics[width=\textwidth,angle=0]{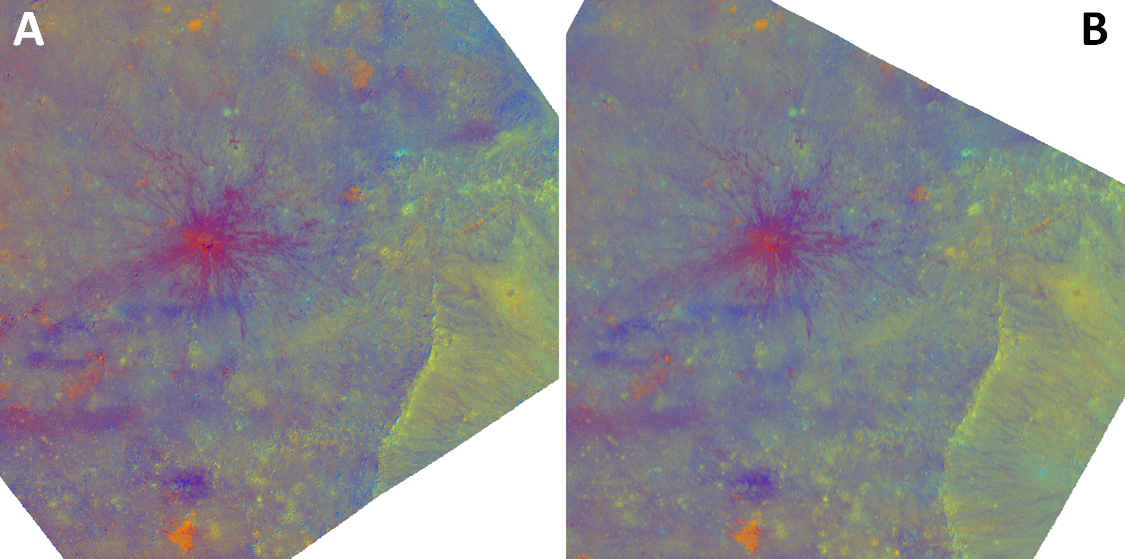}
\caption{Clementine color composites of images of an area around ($14^\circ$, $182^\circ$E), photometrically corrected with the model in Fig.~\ref{fig:phase_curves} and the Akimov disk function. {\bf A}: HAMO images {\bf 8397}, {\bf 8398}, and {\bf 8402} (average phase angle $\bar{\alpha} = 53^\circ$). {\bf B}: HAMO 2 images {\bf 31745}, {\bf 31746}, and {\bf 31750} ($\bar{\alpha} = 31^\circ$). The color scaling is identical for both composites.}
\label{fig:Clem_mosaic_phot}
\end{figure}

The \citet{S13b} photometric model for the clear filter, a fourth-order polynomial, was constructed through a different method, and may be more accurate than the parabola model presented here. Nevertheless, we here provide the clear filter parabola model for reasons of consistency. Figure~\ref{fig:phase_curves} shows that the models agree well. They are very similar for small phase angles, different by about 4\% at intermediate phase angles, and up to 6\% different at high phase angles. In light of the observation bias at high phase angles, we attribute the difference in this range to inaccuracies in the clear filter parabola model, which probably also affect the narrow band filter models. The quality of our photometric models can be further evaluated by comparing color composites of the same area acquired at different phase angles. Figure~\ref{fig:Clem_mosaic_phot} provides such an example for an area west of Marcia crater, which was observed both during {\it HAMO} and {\it HAMO2} at the average phase angle of $53^\circ$ and $31^\circ$, respectively. Despite the $22^\circ$ phase angle difference, the colors of both composites, displayed in identical fashion, agree well.

\subsection{Phase reddening}

\begin{figure}
\centering
\includegraphics[width=8cm,angle=0]{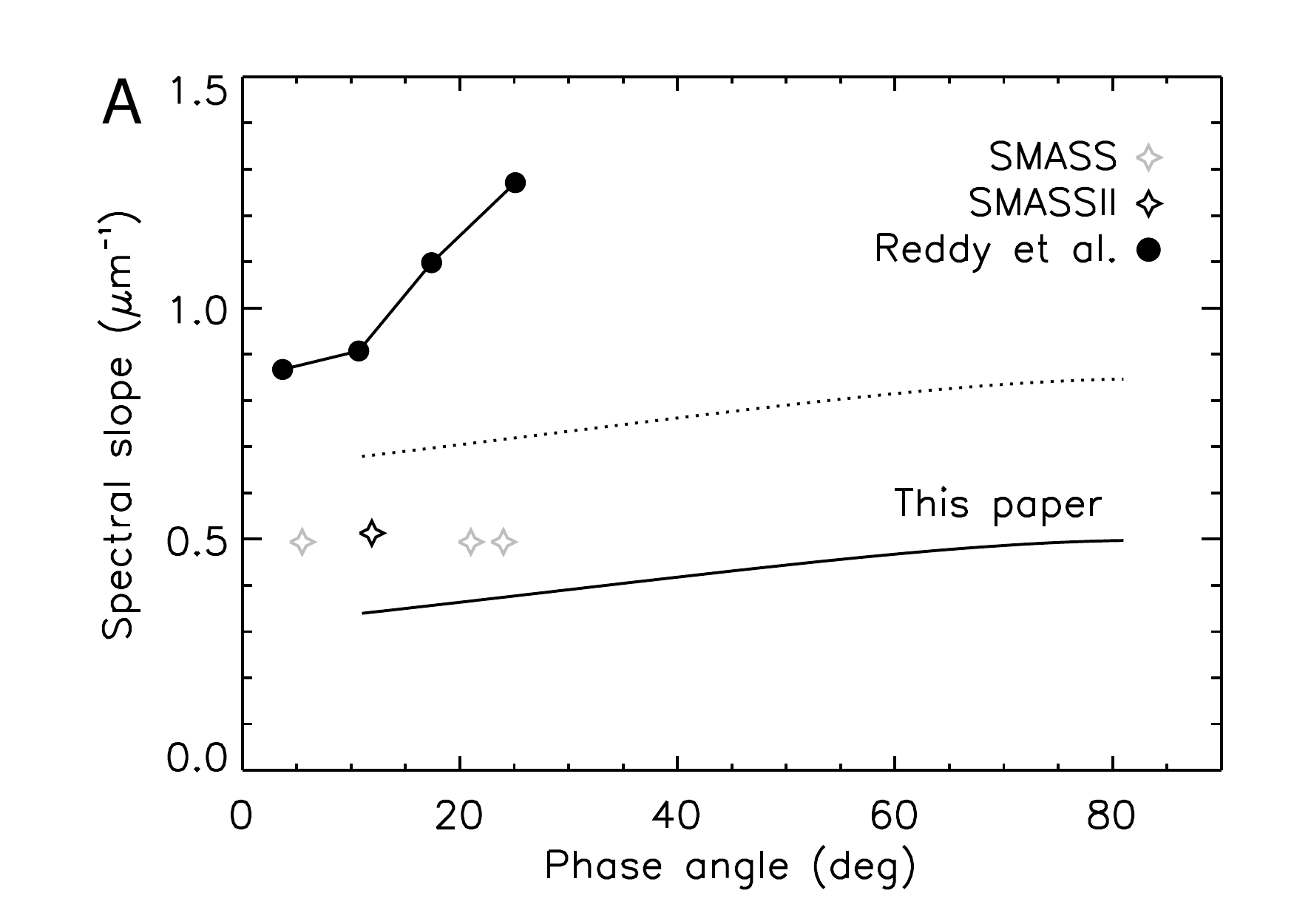}
\includegraphics[width=8cm,angle=0]{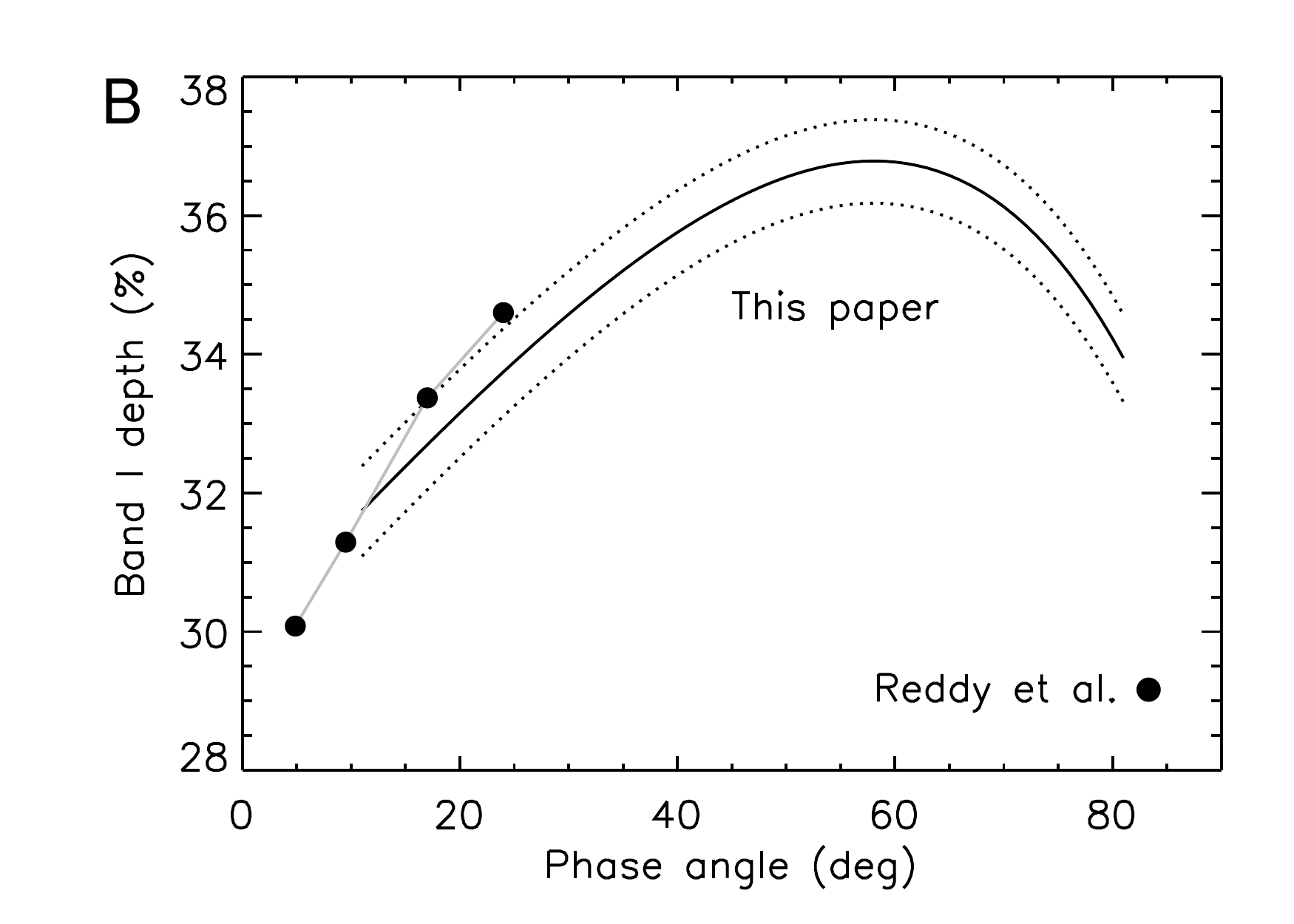}
\caption{Changes in the Vesta spectrum observed with increasing phase angle. {\bf A}. Phase reddening as observed by the FC (drawn line) compared to that observed by \citet{X95} (SMASS), \citet{BB02} (SMASSII), and \citet{R12b}. The spectral slope was calculated over the 0.55-0.75~\textmu m range, except for the \citeauthor{R12b} data (0.55-0.70~\textmu m). The dotted line was calculated assuming that the F2 (0.55~\textmu m) and F3 (0.75~\textmu m) reflectances were over- and underestimated by 3\%, respectively. The SMASS phase angle is one of three shown. {\bf B}. Band~I depth as observed by the FC compared to that observed by \citet{R12b}. The FC band center reflectance was estimated as that for F4 (0.92~\textmu m), and the continuum at the band center was estimated as the F3 reflectance (0.75~\textmu m) multiplied with $1.05 \pm 0.01$ (drawn and dotted lines).}
\label{fig:phase_reddening}
\end{figure}

Like those of other atmosphereless solar system bodies, the spectrum of Vesta has been observed to change with solar phase angle. Often, the slope of the visible spectrum is observed to become steeper with increasing phase (``phase reddening''), but the depth of absorption bands is also known to be variable. \citet{R12b}, hereafter RSN, found such spectral changes for Vesta in ground-based observations. In Fig.~\ref{fig:phase_reddening} we compare their results with the color changes as derived from the phase curves in Fig.~\ref{fig:phase_curves}. The visible spectral slope is defined as $(r_{\lambda_{\rm hi}} - r_{\lambda_{\rm lo}}) / (r_{\lambda_{\rm lo}} (\lambda_{\rm hi} - \lambda_{\rm lo}))$, with $r_\lambda$ being the reflectance at wavelength $\lambda$, $\lambda_{\rm lo} = 0.55$~\textmu m (filter F2), and $\lambda_{\rm hi} = 0.75$~\textmu m (filter F3). The depth of band~I (the 1~\textmu m pyroxene band) is defined as $1 - r_{\rm cen} / r_{\rm con}$, with $r_{\rm cen}$ and $r_{\rm con}$ the reflectances at the band center and continuum, respectively (both center and continuum are defined at the band minimum). Because of the limited wavelength range of the FC we adopt the reflectance at 0.92~\textmu m (filter F4) as the band center reflectance, and estimate the continuum reflectance at the band center from the reflectance at 0.75~\textmu m (filter F3; the blue shoulder). This introduces an uncertainty, as indicated in the figure. In the phase reddening figure we also include the SMASS slope \citep{X95} (Fig.~\ref{fig:comparison_RC1_C0}) and the SMASSII slope \citep{BB02}. Unfortunately, the acquisition date of the SMASS spectrum archived in the NASA Planetary Data System is not documented, and the phase angle can be any of the three values shown in Fig.~\ref{fig:phase_reddening}A.

At small values, the spectral slope is very sensitive to small changes in reflectance. The slope derived from the FC data essentially doubles when we assume that the reflectances in the 0.55~\textmu m and 0.75~\textmu m filters are too high and too low, respectively, by only 3\% (dotted line in Fig.~\ref{fig:phase_reddening}A). In Sec.~\ref{sec:rad_cal} we argued that the ``spectrum'' reconstructed from stray light-corrected FC narrow-band images is consistent with the SMASS spectrum, and thus the spectral slopes are also consistent. The SMASSII slope is very similar to that of SMASS, which implies a small degree of phase reddening (whichever is the true SMASS phase angle), which agrees well with our results. The RSN slopes for the two lowest phase angles ($4^\circ$ and $11^\circ$) are also broadly consistent with the FC and SMASS(II) results, even though the FC 0.55 and 0.75~\textmu m reflectances would need to be off by $+5$\% and $-5$\%, respectively, to match them. However, the increase in the RSN slope for the two highest phase angles ($17^\circ$ and $25^\circ$) is too strong to be consistent with FC/SMASS(II). The discontinuity in their data is also unexpected, given that fit error of the individual slopes is likely small\footnote{The 0.043~\textmu m$^{-1}$ error quoted by the authors was calculated {\it a posteriori} from the scattering of the four spectral slopes around the best-fit line, and does not represent the fitting error of the individual slope values.}. Vesta is unresolved in the RSN spectra, in other words, the spectra are averages over terrains with all possible combinations of the photometric angles. So, if this inconsistency reflects a physical phenomenon, the slope needs to depend on incidence and emission angle and change from disk center to limb. The FC did not observe such behavior; global color images were acquired on approach to Vesta and did not show the strong color gradient across the disk that would follow from the RSN linear trend if reddening depends on the photometric angles. As such, we suspect that the two high-phase angle RSN spectra are affected by calibration errors. The RSN spectra were acquired on different nights, requiring repeated observations of reference and standard stars. On the other hand, the FC image calibration was the same throughout the Vesta campaign. This means that even though the exact value of the slope may be uncertain, slope changes are retrieved relatively reliably. We conclude that the Vesta visible spectral slope changes little with increasing phase angle, confirming \citet{L13}. It is likely in the 0-1~\textmu m$^{-1}$ range below $80^\circ$ phase angle. Phase reddening is weak and well characterized by the FC.

The RSN band depths appear to be consistent with ours (Fig.~\ref{fig:phase_reddening}B). The FC band depth increases up to a certain phase angle and decreases beyond. However, as the bright south pole was not observed at high phase angles, we do not consider the results in Fig.~\ref{fig:phase_reddening} very reliable beyond $60^\circ$. With this in mind, the FC band depth changes are similar to that observed by the VIR spectrometer \citep{Lo13}, which is much better suited for determining this kind of spectral changes.

\section{Conclusions}
\label{sec:conclusions}

We present a method for removal of in-field stray light that provides a first order correction to images of the surface of Vesta. Our method is easy to implement, at negligible computational cost. In addition, we improved the laboratory flat fields using images acquired at Vesta. The radiometric calibration of point sources by \citet{S13a} is good to a few percent for all filters. With the method described in this paper we achieve the same accuracy for images of extended sources. The reflectance spectra reconstructed from the narrow band filter images can serve as a reference for the calibration of the VIS channel of the VIR spectrometer \citep{dS11}. In principle, the stray-light pattern depends on the scene, and is unique to each image. However, for images in which the brightness is more or less uniform, our average stray light patterns are a good first-order approximation. Images that satisfy this criterion are the majority taken at Vesta, mostly as part of the {\it High} and {\it Low Altitude Mapping Orbits} ({\it HAMO}/{\it LAMO}). For images that have the limb in the field of view or that otherwise show brightness gradients, a successful stray light correction is not guaranteed. Calibrated images can be photometrically corrected using our models to create nearly seamless mosaics. The revised flat fields and stray light patterns are delivered to the Dawn Public Data archive, including a bad pixel list. Mosaics constructed from images calibrated with the method in this paper in combination with \citet{S13a} are delivered to the Dawn archive and incorporated in JVesta, the Vesta version of the JMARS geographic information system.

\section*{Acknowledgements}

The authors thank Jason Soderblom and an anonymous referee for their valuable comments.



\bibliography{straylight}

\newpage
\clearpage

\begin{table}
\centering
\caption{Stray light fraction $f$ as estimated from radiance profiles (positive/negative diagonals, row 500) of {\it Survey} images {\bf 3823}-{\bf 3829} ({\it C0}) and {\bf 4054}-{\bf 4060} ({\it C1}). The mean values are adopted as $f^i$ in Eq.~\ref{eq:stray_light} ($f^1 = 0$). See Fig.~\ref{fig:stray_light_estimate} for an illustration of the method.}
\vspace{5mm}
\begin{tabular}{llllll}
\hline
Filter & {\it C0}& {\it C0}& {\it C0} & {\it C1}& Mean \\
       & $+$diag & $-$diag & row 500  & $-$diag & ($f^i$) \\
\hline
F2 & 0.06-0.07 & 0.06 & 0.06 & 0.06-0.07 & 0.06 \\
F3 & 0.04 & 0.04 & 0.04 & 0.06 & 0.05 \\
F4 & 0.09-0.10 & 0.09-0.11 & 0.08 & 0.11 & 0.10 \\
F5 & 0.04-0.06 & 0.04-0.06 & 0.05 & 0.05-0.07 & 0.05 \\
F6 & 0.11-0.14 & 0.11-0.14 & 0.11 & 0.13-0.14 & 0.12 \\
F7 & 0.10 & 0.10 & 0.10 & 0.11 & 0.10 \\
F8 & 0.10 & 0.09 & 0.09 & 0.11 & 0.10 \\
\hline
\end{tabular}
\label{tab:stray_light_faction}
\end{table}

\begin{table}
\centering
\caption{FC2 narrow band filter responsivity factors in units of [$10^6$ J$^{-1}$ m$^2$ nm sr] for targets with a solar and Vesta spectrum. Note that the $\mathcal{R}^2$, $\mathcal{R}^6$, and $\mathcal{R}^7$ values listed in Table~3 of \citet{S13a} were incorrect.}
\vspace{5mm}
\begin{tabular}{lcccccccl}
\hline
Target & $\mathcal{R}^2$ & $\mathcal{R}^3$ & $\mathcal{R}^4$ & $\mathcal{R}^5$ & $\mathcal{R}^6$ & $\mathcal{R}^7$ & $\mathcal{R}^8$ & Source \\
\hline
Solar & 1.93 & 3.85 & 1.82 & 1.76 & 2.47 & 3.22 & 0.218 & \citet{S13a} \\
Vesta & 1.93 & 3.85 & 1.82 & 1.72 & 2.47 & 3.22 & 0.221 & This paper \\
\hline
\end{tabular}
\label{tab:responsivities}
\end{table}

\begin{table}
\centering
\caption{Photometric model coefficients for a phase curve of the form $A_{\rm eq}(\alpha) = a + b \alpha + c \alpha^2$, with phase angle $\alpha$ in degrees. $A_{\rm eq}$ is the equigonal albedo associated with the Akimov disk function (Eq.~\ref{eq:eq_albedo}, Fig.~\ref{fig:phase_curves}). These are model coefficients for the "Solar" responsivities in Table~\ref{tab:responsivities}.}
\vspace{5mm}
\begin{tabular}{lccc}
\hline
Filter & $a$ & $b$ & $c$ \\
\hline
F1      & 0.275 &   $-0.00319$ &  $1.209 \cdot 10^{-5}$ \\
F2      & 0.266 &   $-0.00279$ &  $0.863 \cdot 10^{-5}$ \\
F3      & 0.283 &   $-0.00283$ &  $0.808 \cdot 10^{-5}$ \\
F4      & 0.208 &   $-0.00258$ &  $1.139 \cdot 10^{-5}$ \\
F5      & 0.212 &   $-0.00248$ &  $1.005 \cdot 10^{-5}$ \\
F6      & 0.250 &   $-0.00279$ &  $1.022 \cdot 10^{-5}$ \\
F7      & 0.267 &   $-0.00267$ &  $0.733 \cdot 10^{-5}$ \\
F8      & 0.241 &   $-0.00271$ &  $0.941 \cdot 10^{-5}$ \\
\hline
\end{tabular}
\label{tab:phot_mod_coef}
\end{table}

\newpage
\clearpage

\end{document}